\documentclass[aps,prd,twocolumn,superscriptaddress,nofootinbib,floatfix,10pt]{revtex4-2}

\usepackage{amsmath,amsfonts,amssymb}
\usepackage[utf8]{inputenc}
\usepackage{orcidlink}
\usepackage{xcolor}
\usepackage{hyperref}
\hypersetup{hidelinks}
\hypersetup{colorlinks=true,linkcolor=blue,citecolor=blue,urlcolor=blue}

\makeatletter
\g@addto@macro\normalsize{%
\setlength{\abovedisplayskip}{10pt plus 2pt minus 2pt}%
\setlength{\belowdisplayskip}{10pt plus 2pt minus 2pt}%
\setlength{\abovedisplayshortskip}{6pt plus 2pt}%
\setlength{\belowdisplayshortskip}{8pt plus 2pt minus 2pt}%
}
\makeatother

\newcommand{\dd}{\mathrm{d}}
\newcommand{\MPl}{M_{\rm Pl}}
\newcommand{\cA}{\mathcal{A}}
\newcommand{\lz}{l_0}
\newcommand{\Leff}{\Lambda_{\rm eff}}
\newcommand{\LdS}{L}
\newcommand{\pperp}{p_{\perp}}

\begin{document}

\title{Geodesic completeness and photon propagation\\ in a superconducting T-duality black hole}

\author{Kimet Jusufi\,\orcidlink{0000-0003-0527-4177}}
\email{kimet.jusufi@unite.edu.mk}
\affiliation{Physics Department, University of Tetova, 1200 Tetova, North Macedonia}

\date{\today}

\begin{abstract}
The purpose of this article is twofold. First, we study the geodesic completeness of the string T-duality black hole with the Raychaudhuri equation, for radial massive particles of arbitrary specific energy $\varepsilon$. Exact proper-time quadratures reveal three classes of observers: supercritical observers reach the regular centre in finite proper time through a caustic and pass through it, critical observers ($\varepsilon=1$) approach it only asymptotically with finite expansion, and subcritical observers are reflected at a fold caustic. Because the lapse is analytic and even in $r$, geodesics continue regularly through the centre, and all three classes are complete on the stipulated maximal extension. A diverging expansion therefore signals a caustic of the congruence rather than an endpoint of the geodesics, in contrast with Schwarzschild spacetime. Second, radial Maxwell rays are likewise complete but form a single class, since a massless ray carries no energy scale. We propose to model the black-hole core itself as a superconductor, with London response $\mu^2=4\pi\kappa\rho$ attributed to the existing source. A symmetry-breaking potential with a vacuum-energy offset reproduces the de Sitter core and its photon mass exactly at the centre and approximately in the inner core; since no static scalar theory reproduces the anisotropic source further out, the full radial response remains phenomenological. The resulting Hamiltonian rays follow timelike geodesics of $\mu^2g_{\mu\nu}$ up to reparametrisation and reproduce the three classes, separated by the central mass $\mu_0$, while the density gradient of the core adds a defocusing contribution near the centre. We also distinguish Hamiltonian completeness from completeness of the conformal metric.
\end{abstract}

\maketitle

\section{Introduction}
\label{sec:intro}

The singularity theorems of Penrose and Hawking~\cite{Penrose:1965,Hawking:1970,HawkingEllis} establish geodesic incompleteness under appropriate causal and energy assumptions. Regular black holes replace the central curvature singularity with a finite-density core, often approximately de Sitter~\cite{Bardeen:1968,Dymnikova:1992,AyonBeato:2000,Hayward:2006,Nicolini:2006}. Finite curvature alone, however, does not establish completeness: geodesics must continue through the centre and across horizons on a specified extension~\cite{HawkingEllis,Zhou:2022}. Violating the strong energy condition is also insufficient to evade Penrose's trapped-surface theorem. In stationary regular models satisfying the null energy condition, an inner Cauchy horizon can obstruct global hyperbolicity~\cite{Borissova:2025}, while its susceptibility to mass inflation remains a separate dynamical question~\cite{Poisson:1990,CarballoRubio:2018}.

A concrete setting is the string T-duality black hole of Nicolini, Spallucci and Wondrak~\cite{Nicolini:2019}, based on a propagator with zero-point length $\lz$~\cite{Padmanabhan:1997,Smailagic:2003,Fontanini:2006}. Its lapse has Bardeen's form, with $\lz$ replacing the magnetic charge, without requiring nonlinear electrodynamics; charged extensions have also been investigated~\cite{Gaete:2022ukm,Chen:2026nna}. Related applications of zero-point length include regularisation of naked-singularity interiors~\cite{JusufiLobo:2026dressing}, remnant dynamics in neutron-star matter~\cite{JusufiLobo:2026neutron}, and topological arguments for black-hole regularity~\cite{AnandJusufiBambi:2026topology}. These settings motivate a careful distinction between geometric regularity, geodesic completeness and the additional assumptions needed for a physical matter model.

Using the Raychaudhuri equation~\cite{Raychaudhuri:1955}, Jusufi and Nicolini~\cite{Jusufi:2024} studied radial observers released from rest at infinity, with specific energy $\varepsilon=1$, and found that their expansion remains finite and that they approach the centre only at infinite proper time. We revisit this calculation and generalise it to arbitrary $\varepsilon$, keeping the variation of the Raychaudhuri coefficients along each trajectory. Exact proper-time quadratures, together with closed forms in the de Sitter core, distinguish crossing, critical and reflected observers. Only the critical family approaches the centre asymptotically; supercritical observers cross it in finite proper time through a caustic. The analyticity and parity of the lapse guarantee a unique, regular continuation through the centre. Completeness therefore does not require a bounded expansion, and the difference from Schwarzschild spacetime lies in the continuation of individual geodesics rather than in the absence of focusing.

Radial Maxwell rays provide a complementary test. Their expansion is $\theta=-2E/r$, and they pass through the centre despite its divergence, so they are complete. They form, however, a single class: a massless ray carries no energy scale, so the core can neither screen it from the centre nor hold it there. The Ricci term in the vector-potential wave equation, which takes the form of a mass term in the de Sitter core~\cite{Jusufi:2024}, does not supply such a scale: Maxwell theory is conformally invariant in four dimensions, the Ricci term drops out of geometric optics, and the rays remain null. To introduce a genuine dispersive scale, we prescribe a neutral superconducting response, motivated by London electrodynamics and the Anderson--Higgs mechanism~\cite{London:1935,Ginzburg:1950,Weinberg:1986,Anderson:1963,Higgs:1964}. As in plasma-ray models~\cite{Synge:1960,Perlick:2017,Perlick:2022} and covariant London descriptions~\cite{Bhattacharjee:2019,Turimov:2026}, the response is prescribed: we assume that the existing source itself has the London response $\mu^2(r)=4\pi\kappa\rho(r)$, with zero background field and current, and no separate carrier cloud.

The central London mass satisfies $\mu_0^2=\kappa\Leff/2$. The associated Hamiltonian rays exhibit the same three classes as massive observers, separated by the threshold $E=\mu_0$, and their focusing equation acquires an additional density-gradient term. Since the background stress is already included in the metric, only the quadratic electromagnetic response is added. A homogeneous charged-field condensate whose potential energy supplies the de Sitter core realises this picture exactly at the centre and approximately throughout the inner core, but no static scalar theory reproduces the full radial source. The ray-level results hold for any $\kappa>0$, while their interpretation as localised wave packets near threshold requires $\kappa\gg1$. Completeness refers to radial Hamiltonian trajectories on the stipulated maximal extension and does not establish inner-horizon stability or a universal wave-transmission threshold.

Section~\ref{sec:background} treats massive observers, their continuation through the centre and the contrast with Schwarzschild spacetime; Sec.~\ref{sec:null} analyses Maxwell rays and motivates a genuine photon mass. Sections~\ref{sec:mass}--\ref{sec:waves} introduce the response model, consistency conditions and wave equations, and Secs.~\ref{sec:rays} and~\ref{sec:complete} study Hamiltonian rays and their completeness. We use $G=c=1$, Gaussian units and signature $(-,+,+,+)$, restoring $\hbar$ and $G$ where needed.

\section{Geodesic completeness of the T-duality black hole}
\label{sec:background}

\subsection{Geometry}
\label{sec:geometry}

T-duality replaces the Newtonian potential $-M/r$ by $-M/\sqrt{r^{2}+\lz^{2}}$, with $\lz=2\pi\sqrt{\alpha'}$. The resulting static metric is~\cite{Nicolini:2019}
\begin{equation}
\dd s^{2} = -f(r)\,\dd t^{2} + \frac{\dd r^{2}}{f(r)} + r^{2}\dd\Omega^{2},
\label{eq:metric}
\end{equation}
with lapse
\begin{equation}
f(r) = 1 - \frac{2Mr^{2}}{\left(r^{2}+\lz^{2}\right)^{3/2}} .
\label{eq:f}
\end{equation}
Einstein's equations assign it an anisotropic source with $p_r=-\rho$, energy density
\begin{equation}
\rho = \frac{3 M \lz^{2}}{4\pi\left(r^{2}+\lz^{2}\right)^{5/2}},
\label{eq:rho}
\end{equation}
and tangential pressure
\begin{equation}
\pperp = \rho\,\frac{3r^{2}-2\lz^{2}}{2\left(r^{2}+\lz^{2}\right)} .
\label{eq:pperp}
\end{equation}
The density is finite at the centre, $\rho_0=3M/(4\pi\lz^{3})$, and falls off as $r^{-5}$ outside the core. Two horizons $r_-<r_+$ exist for $M>M_{\rm ext}=3\sqrt{3}\,\lz/4$ and merge at $r_{\rm ext}=\sqrt{2}\,\lz$. The metric carries no electric charge and serves as the fixed background throughout.

Near the centre the geometry is de Sitter~\cite{Nicolini:2019},
\begin{equation}
f \simeq 1 - \frac{r^{2}}{\LdS^{2}}, \qquad \LdS^{2} = \frac{\lz^{3}}{2M},
\label{eq:dS}
\end{equation}
with $R_{\mu\nu}\simeq\Leff g_{\mu\nu}$ and effective cosmological constant
\begin{equation}
\Leff = \frac{3}{\LdS^{2}} = \frac{6M}{\lz^{3}}.
\label{eq:Leff}
\end{equation}
For $M\gg\lz$ one has $r_+\simeq2M-3\lz^{2}/4M$ and $r_-\simeq\LdS(1+3\lz/8M)$. The spacetime then consists of three zones: a static, nearly de Sitter core $r<r_-$; a dynamical shell $r_-<r<r_+$ in which $r$ is timelike; and a nearly Schwarzschild exterior. The null energy condition holds everywhere, since $\rho+\pperp=5r^{2}\rho/[2(r^{2}+\lz^{2})]\geq0$, whereas the strong energy condition fails for $r<\sqrt{2/3}\,\lz$~\cite{Jusufi:2024}. Two properties of $f$ control everything that follows: $f(0)=1$ is its maximum, with $f\to1$ as $r\to\infty$, and $f$ is a smooth function of $r^{2}$, so the centre is an ordinary point of the manifold.

\subsection{Radial geodesics and the Raychaudhuri equation}
\label{sec:raych}

We now revisit and generalise the massive-particle analysis of Ref.~\cite{Jusufi:2024}, extending radial infall from rest at infinity ($\varepsilon=1$) to arbitrary positive specific energy $\varepsilon$. We retain the position dependence of the Raychaudhuri coefficients and distinguish a caustic of the congruence from an endpoint of an individual geodesic. For later use, we first state the radial kinematic identities shared by massive observers, Maxwell rays and the massive-photon Hamiltonian rays of Sec.~\ref{sec:complete}. Let an ingoing radial congruence with parameter $s$ obey
\begin{equation}
\left(\frac{\dd r}{\dd s}\right)^{2}=\mathcal{E}^{2}-U(r)\equiv P^{2} ,
\label{eq:Pdef}
\end{equation}
with tangent $(\mathcal{E}/f,-P,0,0)$ in static coordinates. Since $\sqrt{-g}\propto r^{2}$ is independent of $f$, its expansion is
\begin{equation}
\theta = -\frac{1}{r^{2}}\frac{\dd}{\dd r}\!\left(r^{2}P\right) = -\frac{2P}{r} + \frac{U'}{2P} ,
\label{eq:master-theta}
\end{equation}
and its rate of change along the flow, $\dd/\dd s=-P\,\dd/\dd r$, is
\begin{equation}
\frac{\dd\theta}{\dd s} = -\frac{2P^{2}}{r^{2}} - \frac{U'}{r} - \frac{U''}{2} - \frac{U'^{2}}{4P^{2}} .
\label{eq:master}
\end{equation}
Both are scalar functions of $r$ and remain valid across the horizons, where the static chart fails but regular ingoing or outgoing coordinates can be used.

For a massive observer the parameter is proper time, $s=\tau$, the potential is $U=f$, and $\mathcal{E}=\varepsilon=-u_t$ is the specific energy. The four-velocity is
\begin{equation}
u^{\mu}=\left(\frac{\varepsilon}{f},\,-P_\varepsilon,\,0,\,0\right), \qquad P_\varepsilon=\sqrt{\varepsilon^{2}-f} ,
\label{eq:u-eps}
\end{equation}
and Eqs.~\eqref{eq:master-theta} and~\eqref{eq:master} give
\begin{equation}
\Theta_\varepsilon=-\frac{2P_\varepsilon}{r}+\frac{f'}{2P_\varepsilon} ,
\label{eq:Theta-eps}
\end{equation}
and its rate of change along the flow,
\begin{equation}
\frac{\dd\Theta_\varepsilon}{\dd\tau}=-\frac{2P_\varepsilon^{2}}{r^{2}}-\frac{f'}{r}-\frac{f''}{2}-\frac{f'^{2}}{4P_\varepsilon^{2}} .
\label{eq:dTheta-eps}
\end{equation}
These satisfy, identically in $\varepsilon$, the Raychaudhuri equation for a twist-free timelike congruence~\cite{Raychaudhuri:1955,HawkingEllis},
\begin{equation}
\frac{\dd\Theta}{\dd\tau} = -\frac{1}{3}\Theta^{2} - \sigma_{\mu\nu}\sigma^{\mu\nu} - R_{\mu\nu}u^{\mu}u^{\nu} ,
\label{eq:ray-JN}
\end{equation}
with shear $\sigma_{\mu\nu}\sigma^{\mu\nu}=\tfrac{2}{3}\left(P_\varepsilon/r-P_\varepsilon'\right)^{2}$ and focusing term
\begin{equation}
R_{\mu\nu}u^{\mu}u^{\nu} = \frac{3M\lz^{2}\left(3r^{2}-2\lz^{2}\right)}{\left(r^{2}+\lz^{2}\right)^{7/2}} .
\label{eq:Ruu}
\end{equation}
The focusing term is independent of $\varepsilon$, because $R^{t}{}_{t}=R^{r}{}_{r}$ implies $R_{\mu\nu}u^{\mu}u^{\nu}=-R^{t}{}_{t}$ for every radial unit timelike vector; it is negative, i.e.\ defocusing, where the strong energy condition fails.

The observers of Ref.~\cite{Jusufi:2024}, which fall from rest at infinity, have $\varepsilon=1$. Their expansion is
\begin{equation}
\Theta_1 = -\frac{3\sqrt{2M}\left(r^{2}+2\lz^{2}\right)}{2\left(r^{2}+\lz^{2}\right)^{7/4}} ,
\label{eq:Theta-JN}
\end{equation}
while the rate of change of the expansion is
\begin{equation}
\frac{\dd\Theta_1}{\dd\tau} = -\frac{9Mr^{2}\left(r^{2}+\tfrac{10}{3}\lz^{2}\right)}{2\left(r^{2}+\lz^{2}\right)^{7/2}} .
\label{eq:dTheta-JN}
\end{equation}
The expansion is finite everywhere, with $\Theta_1(0)=-3/\LdS$, and its rate vanishes at the centre, because the de Sitter defocusing $-R_{\mu\nu}u^{\mu}u^{\nu}\to\Leff$ cancels $-\Theta_1^{2}/3\to-\Leff$. Since $f(0)=1$ is the maximum of $f$, these observers have exactly the energy needed to reach the centre. They are the critical members of a one-parameter family, and the question is which of their properties survive for $\varepsilon\neq1$.

\subsection{Proper time from the Raychaudhuri equation}
\label{sec:tau}

Following Ref.~\cite{Jusufi:2024}, we divide Eq.~\eqref{eq:ray-JN} by $\Theta_\varepsilon^{2}$ and write the left-hand side as a derivative of $1/\Theta_\varepsilon$:
\begin{equation}
\frac{\dd}{\dd\tau}\!\left(\frac{1}{\Theta_\varepsilon}\right)=\frac{1}{3}+K_\varepsilon\equiv D_\varepsilon ,
\label{eq:inv-theta}
\end{equation}
where $K_\varepsilon=(\sigma_{\mu\nu}\sigma^{\mu\nu}+R_{\mu\nu}u^{\mu}u^{\nu})/\Theta_\varepsilon^{2}$. All quantities can be expressed through two functions,
\begin{align}
A_\varepsilon &= 2(\varepsilon^{2}-1)\left(r^{2}+\lz^{2}\right)^{5/2} \nonumber\\
&\quad+3Mr^{2}\left(r^{2}+2\lz^{2}\right), \label{eq:Aeps}\\
B_\varepsilon &= 3M^{2}r^{6}\left(3r^{2}+10\lz^{2}\right)+2(\varepsilon^{2}-1)^{2}\left(r^{2}+\lz^{2}\right)^{5} \nonumber\\
&\quad+M(\varepsilon^{2}-1)\,r^{2}\left(r^{2}+\lz^{2}\right)^{3/2} \nonumber\\
&\quad\times\left(2\lz^{4}+25\lz^{2}r^{2}+8r^{4}\right), \label{eq:Beps}
\end{align}
in terms of which, exactly,
\begin{equation}
\frac{1}{\Theta_\varepsilon}=-\frac{r\left(r^{2}+\lz^{2}\right)^{5/2}P_\varepsilon}{A_\varepsilon}, \qquad
D_\varepsilon=\frac{B_\varepsilon}{A_\varepsilon^{2}} .
\label{eq:invTheta-eps}
\end{equation}
For $\varepsilon=1$ these functions reduce to
\begin{equation}
A_1=3Mr^{2}\left(r^{2}+2\lz^{2}\right),
\end{equation}
along with
\begin{equation}
B_1=3M^{2}r^{6}\left(3r^{2}+10\lz^{2}\right),
\end{equation}
so that
\begin{equation}
D_1=\frac{r^{2}\left(3r^{2}+10\lz^{2}\right)}{3\left(r^{2}+2\lz^{2}\right)^{2}} .
\end{equation}
Along an infalling observer, $r=r(\tau)$ and $\dd r/\dd\tau=-P_\varepsilon$. Therefore the exact integrated form of Eq.~\eqref{eq:inv-theta}, with $\tau=0$ at $r=r_0$ and $\Theta_\varepsilon=\Theta_0$, is
\begin{equation}
\frac{1}{\Theta_\varepsilon(\tau)}-\frac{1}{\Theta_0}
=\int_0^\tau D_\varepsilon(r(s))\,\dd s .
\label{eq:inverse-time-integral}
\end{equation}
The coefficient $D_\varepsilon$ must be evaluated along the trajectory; replacing it by a constant is only a local approximation and is not used below.

On a branch where $\Theta_\varepsilon\neq0$ and $D_\varepsilon\neq0$, the inverse expansion can instead be used as a trajectory variable:
\begin{equation}
\tau=\int_{1/\Theta_0}^{1/\Theta_\varepsilon}
\frac{\dd (\Theta_\varepsilon^{-1})}{D_\varepsilon}.
\label{eq:tau-int}
\end{equation}
This change of variable must be applied piecewise if $\Theta_\varepsilon$ vanishes or $D_\varepsilon$ changes sign. For $\varepsilon\geq1$, the expressions above give $A_\varepsilon>0$ and $B_\varepsilon>0$ at finite $r>0$, so $\Theta_\varepsilon<0$ and $X=\Theta_\varepsilon^{-1}$ is monotonic along the infall. For subcritical observers, $B_\varepsilon$ can be negative and no global monotonicity is assumed. The time-domain relation~\eqref{eq:inverse-time-integral} applies on intervals with $\Theta_\varepsilon\neq0$ without requiring $D_\varepsilon\neq0$; the radial quadrature below avoids both divisions.

Using the radial geodesic equation one can also write an exact proper time relation on each ingoing branch as
\begin{equation}
\tau(r)=\int_r^{r_0}\frac{\dd x}{P_\varepsilon(x)}
=\int_r^{r_0}\frac{\dd x}{\sqrt{\varepsilon^2-f(x)}} .
\label{eq:JN21-eps}
\end{equation}
The outgoing branch is treated with the opposite sign of $\dd r/\dd\tau$. Equation~\eqref{eq:JN21-eps} determines the elapsed proper time without an assumption about the variation of $K_\varepsilon$.

For critical observers, $P_1^{2}=1-f=2Mr^{2}/(r^{2}+\lz^{2})^{3/2}$, and the exact integrand is particularly simple:
\begin{equation}
\tau(r)=\frac{1}{\sqrt{2M}}\int_r^{r_0}
\frac{(x^2+\lz^2)^{3/4}}{x}\,\dd x .
\label{eq:tau-critical-exact}
\end{equation}
This integral can be evaluated in closed form. With the variable
\begin{equation}
w(r)=\left(1+\frac{r^{2}}{\lz^{2}}\right)^{1/4} ,
\label{eq:w-def}
\end{equation}
the integrand becomes $2\LdS\,w^{6}\dd w/(w^{4}-1)$, with $\LdS=\lz^{3/2}/\sqrt{2M}$, and partial fractions then give
\begin{equation}
\tau(r)=\LdS\left[F(w_0)-F\bigl(w(r)\bigr)\right] ,
\label{eq:tau-critical-closed}
\end{equation}
where $w_0=w(r_0)$ and
\begin{equation}
F(w)=\frac{2}{3}\,w^{3}+\frac{1}{2}\ln\frac{w-1}{w+1}+\arctan w .
\label{eq:Fw}
\end{equation}
As $r\to0$, $w-1\simeq r^{2}/(4\lz^{2})$, and the logarithm in $F$ behaves as $\tfrac12\ln(w-1)\simeq\ln r-\ln(2\lz)$. The near-centre behaviour is therefore
\begin{equation}
\tau(r)=\LdS\ln\frac{r_0}{r}+C(r_0)
+O\!\left(\LdS\frac{r^2}{\lz^2}\right)
\longrightarrow+\infty ,
\label{eq:tau-critical-asymptotic}
\end{equation}
with the finite constant
\begin{equation}
C(r_0)=\LdS\left[F(w_0)-\frac{2}{3}-\frac{\pi}{4}-\ln\frac{r_0}{2\sqrt{2}\,\lz}\right] .
\label{eq:C-r0}
\end{equation}
Thus the infinite proper time is logarithmic, and it follows from the exact radial equation without any approximation for $K_\varepsilon$.

The Raychaudhuri route gives the same proper time, and the two agree exactly rather than only asymptotically. By Eq.~\eqref{eq:invTheta-eps}, the inverse expansion $X=1/\Theta_\varepsilon$ is a function of $r$ alone. Its derivative along the flow can therefore be computed in two ways: from the Raychaudhuri equation~\eqref{eq:inv-theta}, $\dd X/\dd\tau=D_\varepsilon$, and from the chain rule, $\dd X/\dd\tau=(\dd X/\dd r)(\dd r/\dd\tau)=-P_\varepsilon\,\dd X/\dd r$. Equating the two gives the identity
\begin{equation}
D_\varepsilon=-P_\varepsilon\,\frac{\dd X}{\dd r} ,
\label{eq:D-identity}
\end{equation}
which can also be verified directly from Eqs.~\eqref{eq:Aeps}--\eqref{eq:invTheta-eps}. Substituting it into Eq.~\eqref{eq:tau-int} gives
\begin{equation}
\tau=\int_{1/\Theta_0}^{1/\Theta_\varepsilon}\frac{\dd X}{D_\varepsilon}
=\int_r^{r_0}\frac{\dd x}{P_\varepsilon(x)} ,
\label{eq:equivalence}
\end{equation}
which is Eq.~\eqref{eq:JN21-eps}. The two approaches therefore give the same proper time on every branch on which $\dd X/\dd r\neq0$, which is the same condition, $D_\varepsilon\neq0$, needed to use $X$ as a trajectory variable. Dividing the Raychaudhuri equation by $\Theta_\varepsilon^{2}$ introduces no approximation; it only replaces the radius by the inverse expansion as the integration variable.

For the critical observers the equivalence can be seen explicitly near the centre. There the inverse expansion tends to a finite value,
\begin{equation}
\frac{1}{\Theta_1}=-\frac{\LdS}{3}-\frac{5\LdS}{12\lz^{2}}\,r^{2}+O\!\left(\LdS\frac{r^{4}}{\lz^{4}}\right) ,
\label{eq:invTheta1-series}
\end{equation}
while the right-hand side of Eq.~\eqref{eq:inv-theta} vanishes quadratically,
\begin{equation}
D_1=\frac{5r^2}{6\lz^2}+O\!\left(\frac{r^4}{\lz^4}\right) .
\label{eq:D1-series}
\end{equation}
With $P_1\simeq r/\LdS$, Eqs.~\eqref{eq:invTheta1-series} and~\eqref{eq:D1-series} satisfy the identity~\eqref{eq:D-identity}, and the integrand of Eq.~\eqref{eq:tau-int} becomes
\begin{equation}
\frac{\dd(\Theta_1^{-1})}{D_1}
=-\frac{\LdS}{r}\left[1+O\!\left(\frac{r^2}{\lz^2}\right)\right]\dd r .
\label{eq:inverse-critical-asymptotic}
\end{equation}
Integrating from $r_0$ to $r$ reproduces the logarithm of Eq.~\eqref{eq:tau-critical-asymptotic}. The infinite proper time thus arises from the ratio of two quantities, the shrinking change of $1/\Theta_1$ and the vanishing of $D_1$, neither of which alone is singular.

For $\varepsilon>1$, by contrast, $P_\varepsilon\to\sqrt{\varepsilon^2-1}>0$ and the integral to the centre is finite. For $0<\varepsilon<1$, the inner simple turning point $r_t$ gives an integrable square-root endpoint, so the proper time to reflection is finite. These statements, evaluated below, concern the time to the centre or turning point; global completeness also requires continuation of the geodesics on the chosen extension.

\subsection{Three classes of observers and geodesic completeness}
\label{sec:classes}

A geodesic is complete if its affine parameter ranges over the whole real line, and incomplete if its maximal affine-parameter interval has a finite endpoint~\cite{HawkingEllis}. Reaching a regular point at finite proper time is therefore no sign of incompleteness, and completeness does not require an infinite proper time to reach the centre. The behaviour near the centre is governed by $D_\varepsilon$ there. With $f\simeq1-r^{2}/\LdS^{2}$ one has $P_\varepsilon^{2}\simeq\varepsilon^{2}-1+r^{2}/\LdS^{2}$, and the sign of $\varepsilon^{2}-1$ decides.

\emph{Critical observers, $\varepsilon=1$.} Here $D_1$ vanishes like $r^{2}$ and $1/\Theta_1$ tends to its finite limit $-\LdS/3$. Keeping the variation of both quantities in Eq.~\eqref{eq:inverse-critical-asymptotic} gives $\dd\tau\sim-\LdS\,\dd r/r$, and hence the logarithmic divergence of Eq.~\eqref{eq:tau-critical-asymptotic}. Near the centre $\Theta_1-\Theta_1(0)\simeq15\sqrt{2M}\,r^{2}/(4\lz^{7/2})$ and Eq.~\eqref{eq:ray-JN} linearises to
\begin{equation}
\frac{\dd\Theta_1}{\dd\tau} \simeq -\frac{2}{\LdS}\left[\Theta_1-\Theta_1(0)\right] ,
\label{eq:Theta-lin}
\end{equation}
whose integral from a point $r_1$ is
\begin{equation}
\tau-\tau_1 \simeq -\frac{\LdS}{2}\,\ln\frac{\Theta_1-\Theta_1(0)}{\Theta_1(r_1)-\Theta_1(0)} = \LdS\,\ln\frac{r_1}{r}\;\longrightarrow\;\infty .
\label{eq:tau-JN}
\end{equation}
The centre is approached only asymptotically, $r\propto e^{-\tau/\LdS}$, with finite expansion, and the geodesics are future complete~\cite{Jusufi:2024}.

\emph{Supercritical observers, $\varepsilon>1$.} The radial velocity tends to $P_0=\sqrt{\varepsilon^{2}-1}>0$, and
\begin{equation}
\Theta_\varepsilon\simeq-\frac{2P_0}{r}-\frac{2r}{\LdS^{2}P_0}, \qquad
\frac{\dd\Theta_\varepsilon}{\dd\tau}\simeq-\frac{2P_0^{2}}{r^{2}}+\frac{1}{\LdS^{2}} .
\label{eq:Theta-super}
\end{equation}
The shear grows like the expansion, $\sigma_{\mu\nu}\sigma^{\mu\nu}\simeq\Theta_\varepsilon^{2}/6$, the focusing term stays finite, and $D_\varepsilon\to1/2$. Integrating Eq.~\eqref{eq:tau-int} order by order in $1/\Theta$, the proper time from $\Theta_0$ to the centre is
\begin{equation}
\tau_c=\frac{2}{|\Theta_0|}+\frac{40M}{3\lz^{3}|\Theta_0|^{3}}+O\!\left(|\Theta_0|^{-5}\right) ,
\label{eq:caustic-tau}
\end{equation}
where $\tau=0$ is chosen at an initial radius sufficiently close to the centre and the expansion is at fixed $\varepsilon>1$. Thus $2/|\Theta_0|$ is a leading local approximation, not the exact proper time from an arbitrary initial radius. Within the leading de Sitter approximation, $f=1-r^2/\LdS^2$, Eq.~\eqref{eq:tau-int} can be evaluated in closed form without freezing $D_\varepsilon$,
\begin{equation}
\tau_c=\LdS\,\operatorname{arcsinh}\sqrt{\frac{\sqrt{T(T-8)}-(T-12)}{2(T-9)}} ,
\label{eq:tau-Theta0}
\end{equation}
with $T\equiv\LdS^{2}\Theta_0^{2}$, valid for $T>9$ within that de Sitter approximation. Here $\Theta_0$ is evaluated consistently in the same approximate geometry; the full metric is governed by Eq.~\eqref{eq:JN21-eps}. To leading order $\Theta_\varepsilon\simeq-2/(\tau_c-\tau)$ and $r\simeq P_0(\tau_c-\tau)$: the observers reach the centre at finite proper time, where the expansion diverges. The divergence is a caustic of the converging congruence, as for a spherical shell of free particles in flat space, and not a singularity: the curvature is finite and, since $f$ depends smoothly on $r^{2}$, the metric and the geodesic equation are analytic at the origin in Cartesian-like coordinates $x^{i}=rn^{i}$ (Sec.~\ref{sec:continuation}). The geodesics therefore pass through the centre, after which $\Theta_\varepsilon\simeq+2/(\tau-\tau_c)$ describes an expanding congruence. They cross the core, the inner horizon and the shell, where $P_\varepsilon^{2}=\varepsilon^{2}-f>0$, and enter a new exterior region. Because $f\geq f_{\min}$, with $f_{\min}<0$ attained between the horizons, $P_\varepsilon\leq(\varepsilon^{2}-f_{\min})^{1/2}$ and
\begin{equation}
\tau(R)-\tau_c=\int_{0}^{R}\frac{\dd r}{P_\varepsilon}\;\geq\;\frac{R}{\sqrt{\varepsilon^{2}-f_{\min}}}\;\xrightarrow{\;R\to\infty\;}\;\infty .
\label{eq:tau-escape}
\end{equation}
Equivalently, far from the hole $D_\varepsilon\to1/2$ again, so $1/\Theta_\varepsilon\simeq\tau/2$ grows without bound and $\Theta_\varepsilon\to0$ only as $\tau\to\infty$. Traced backwards, the observers came from infinity in their initial exterior region, where $\tau\to-\infty$.

\emph{Subcritical observers, $\varepsilon<1$.} These observers are released from rest at $r_{\rm out}>r_+$, where $f(r_{\rm out})=\varepsilon^{2}$, and cannot reach the centre: they are reflected at the radius $r_t$ in the core where $f(r_t)=\varepsilon^{2}$, with $r_t\simeq\LdS\sqrt{1-\varepsilon^{2}}$. There $P_\varepsilon\to0$, the term $f'/(2P_\varepsilon)$ in Eq.~\eqref{eq:Theta-eps} diverges, and $D_\varepsilon\to1$, so Eq.~\eqref{eq:inv-theta} gives $\tau_t-\tau\simeq1/|\Theta_\varepsilon|$. This is a fold caustic, at which the ingoing and outgoing branches of the congruence meet, reached at finite proper time; the individual geodesics are smooth there. After reflection the observers cross the inner horizon and the shell into a new exterior region, rise to $r_{\rm out}$ and fall into the next black hole, as timelike geodesics do in the Reissner--Nordstr\"om interior. Each cycle lasts
\begin{equation}
T_\varepsilon=2\int_{r_t}^{r_{\rm out}}\frac{\dd r}{P_\varepsilon}>0 ,
\label{eq:period}
\end{equation}
which is finite, since the integrand has only integrable square-root singularities at the turning points. On the maximal extension the cycles repeat indefinitely, so the proper time is unbounded in both directions.

In the de Sitter core the proper time from a radius $r_1$ to the centre, or to the turning point, also follows directly from $\dd\tau=-\dd r/P_\varepsilon$:
\begin{equation}
\Delta\tau_\varepsilon\simeq
\begin{cases}
\LdS\,\operatorname{arcsinh}\!\left[\dfrac{r_1}{\LdS\sqrt{\varepsilon^{2}-1}}\right], & \varepsilon>1,\\[2ex]
\LdS\,\ln\dfrac{r_1}{r}\;\longrightarrow\;\infty, & \varepsilon=1,\\[2ex]
\LdS\,\operatorname{arccosh}\!\left[\dfrac{r_1}{\LdS\sqrt{1-\varepsilon^{2}}}\right], & \varepsilon<1 .
\end{cases}
\label{eq:tau-eps}
\end{equation}
As $\varepsilon\to1$ from either side, the elapsed time diverges logarithmically:
\[
\Delta\tau_\varepsilon\simeq\LdS\ln\!\left[
\frac{2r_1}{\LdS\sqrt{|\varepsilon^2-1|}}\right].
\]
This recovers the critical asymptotics of Eq.~\eqref{eq:tau-JN}.

The results are summarised in Table~\ref{tab:observers}. A finite central expansion and an infinite proper time to the centre are properties of the critical family alone; the critical orbit is an unstable equilibrium, and an arbitrarily small change of $\varepsilon$ replaces it by a crossing or a reflection. Completeness, by contrast, holds for every $\varepsilon$. What distinguishes the classes is only where the proper time becomes infinite: at the centre for $\varepsilon=1$, at infinity in a new exterior region for $\varepsilon>1$, and after infinitely many cycles for $\varepsilon<1$.

\begin{table*}[t]
\caption{Radial observers near the centre of the T-duality black hole, compared with Schwarzschild. $D$ is the right-hand side of Eq.~\eqref{eq:inv-theta} at the centre (at $r_t$ for $\varepsilon<1$, at $r\to0$ for Schwarzschild).}
\label{tab:observers}
\begin{ruledtabular}
\begin{tabular}{lcccc}
& $D$ & $\Theta$ & $\tau$ to $r=0$ or $r_t$ & Complete \\
\hline
$\varepsilon>1$ & $1/2$ & $-\infty$ (caustic) & finite & yes \\
$\varepsilon=1$ & $0$ & $-3/\LdS$ & $\infty$ & yes \\
$\varepsilon<1$ & $1$ & $-\infty$ (fold) & finite & yes \\
Schwarzschild & $1$ & $-\infty$ (singular) & finite & no \\
\end{tabular}
\end{ruledtabular}
\end{table*}

\subsection{Continuation through the centre: analyticity and parity}
\label{sec:continuation}

Supercritical observers, and below Maxwell rays and supercritical photons, reach $r=0$ at a finite value of their parameter. Completeness then requires that they can be continued beyond the centre, and that the continuation be unique and regular. Zhou and Modesto showed that this is not guaranteed by finite curvature alone~\cite{Zhou:2022}. A signed radial coordinate provides one representation of continuation through the centre; when the centre is regular, the same trajectory can instead be described with nonnegative areal radius and antipodal angles. Analyticity is a sufficient regularity condition in the argument used here, rather than a necessary condition for local geodesic uniqueness: a nondegenerate $C^2$ metric already has locally Lipschitz Christoffel symbols and a locally unique geodesic for each initial position and tangent. One must distinguish uniqueness of geodesics in a specified spacetime from the choice of an extension of that spacetime. Within the analytic-continuation prescription discussed in Ref.~\cite{Zhou:2022}, the Hayward lapse $f_{\rm H}=1-2Mr^{2}/(r^{3}+2M\ell^{2})$ develops a pole at $r=-(2M\ell^{2})^{1/3}$, where curvature diverges and the continued geodesics end. This is a statement about that analytic extension; nonanalyticity alone does not establish geodesic incompleteness.

The T-duality metric admits a regular analytic centre and a consistent antipodal continuation. Two properties ensure this. The first is analyticity. The lapse~\eqref{eq:f} is a real-analytic function of $r$ on the whole real line, whose only singularities lie at $r=\pm i\lz$, off the real axis, so its continuation to $r<0$ is unique. The second is parity. The lapse depends on $r$ only through $r^{2}$,
\begin{equation}
\begin{split}
&f(-r)=f(r),\\
&f=1-\frac{2M}{\lz^{3}}\,r^{2}+\frac{3M}{\lz^{5}}\,r^{4}-\frac{15M}{4\lz^{7}}\,r^{6}+\dots ,
\end{split}
\label{eq:parity}
\end{equation}
with only even powers, so the continued metric on $r<0$ is identical to the original one. The map $(r,\vartheta,\varphi)\to(-r,\pi-\vartheta,\varphi+\pi)$ is then an isometry, and the region $r<0$ is the antipodal part of the same spacetime, not a new one. The same statement can be made in Cartesian coordinates $x^{i}=rn^{i}$, in which the metric~\eqref{eq:metric} reads
\begin{equation}
\dd s^{2}=-f\,\dd t^{2}+\delta_{ij}\,\dd x^{i}\dd x^{j}+\frac{1-f}{f\,r^{2}}\left(x_i\,\dd x^{i}\right)^{2},
\label{eq:cartesian}
\end{equation}
with
\begin{equation}
\frac{1-f}{r^{2}}=\frac{2M}{\left(r^{2}+\lz^{2}\right)^{3/2}} .
\label{eq:cartesian-coef}
\end{equation}
Every coefficient is an analytic function of $r^{2}=\delta_{ij}x^{i}x^{j}$, and $f(0)=1\neq0$, so the metric is real-analytic in a neighbourhood of the origin, which is an ordinary point of the chart. The geodesic equations there have analytic coefficients, and their solutions exist, are unique and continue through $x^{i}=0$. For the Hayward metric, by contrast, $(1-f_{\rm H})/r^{2}=2M/(r^{3}+2M\ell^{2})$ contains $|\mathbf{x}|^{3}$ and is not real-analytic at the origin. This failure of analyticity must not be confused with failure of the lower differentiability needed for local geodesic uniqueness.

For radial geodesics the parity argument can be made explicit. The radial equation $(\dd r/\dd\tau)^{2}=\varepsilon^{2}-f$ implies $\ddot r=-\tfrac12 f'(r)$, and $f'$ is odd. The solution through the centre at $\tau=\tau_c$ is therefore odd in $\tau-\tau_c$,
\begin{equation}
r(\tau)=-P_0\,(\tau-\tau_c)+O\!\left((\tau-\tau_c)^{3}\right) ,
\label{eq:r-odd}
\end{equation}
and passes smoothly from positive to negative values, i.e.\ through the centre to the antipodal direction. The expansion~\eqref{eq:Theta-eps} correspondingly changes sign, from $-2P_0/r$ on the ingoing branch to $+2P_0/|r|$ on the outgoing one, i.e.\ from $-2/(\tau_c-\tau)$ to $+2/(\tau-\tau_c)$: the caustic is crossed, not an endpoint. The same holds for Maxwell rays, which have $U=0$, and for massive photons, since the potential $f\mu^{2}$ is also even and analytic: $\mu^{2}(-r)=\mu^{2}(r)$, with singularities only at $r=\pm i\lz$. The continuation through the centre is thus unique and regular for every radial congruence in this paper, and the completeness statements below and in Sec.~\ref{sec:complete} rest on it. The same conclusion holds for the Bardeen metric, which has the same functional form with $\lz$ replaced by the magnetic charge, whereas the signed-radius analytic extension of the Hayward metric considered in Ref.~\cite{Zhou:2022} is incomplete.

\subsection{Contrast with Schwarzschild spacetime}
\label{sec:schw}

The same analysis applied to Schwarzschild spacetime, $f=1-2M/r$, shows what the T-duality core changes. For $\varepsilon=1$, $P_1=\sqrt{2M/r}$ and Eqs.~\eqref{eq:master-theta} and~\eqref{eq:master} give $\Theta_1=-\tfrac32\sqrt{2M}\,r^{-3/2}$ and $\dd\Theta_1/\dd\tau=-\tfrac92 M r^{-3}$, so that $D_1=1$ exactly and the Raychaudhuri equation integrates without approximation:
\begin{equation}
\frac{1}{\Theta_1}-\frac{1}{\Theta_0}=\tau
\quad\Longrightarrow\quad
\Theta_1=-\frac{1}{\tau_s-\tau},\qquad \tau_s=\frac{1}{|\Theta_0|} .
\label{eq:Schw}
\end{equation}
For other energies, $P_\varepsilon^2=\varepsilon^2-1+2M/r$ is dominated by $2M/r$ near the centre. Hence $D_\varepsilon\to1$ and the same leading singular behaviour follows. From a sufficiently small radius $r_1$,
\begin{equation}
\Delta\tau\simeq\frac{2r_1^{3/2}}{3\sqrt{2M}},
\label{eq:Schw-time-asymptotic}
\end{equation}
independently of $\varepsilon$ at leading order. Unlike a regular caustic, this endpoint has divergent curvature, $R_{\mu\nu\rho\sigma}R^{\mu\nu\rho\sigma}=48M^2/r^6$, and divergent tidal forces. Inside the Schwarzschild black-hole horizon, $r$ is timelike and $r=0$ is a spacelike singularity; no regular continuation through it exists.

The T-duality centre is instead an ordinary point of a static core. Its metric is analytic in Cartesian coordinates, its curvature is finite, and geodesics that arrive there continue regularly. Repulsion in the core produces the critical and reflected classes, while sufficiently energetic observers cross the centre. The three classes therefore differ in their local radial behaviour, but none terminates at a singular centre on the prescribed maximal extension.

The expansion scalar belongs to a congruence in either spacetime. For the radial families studied here, its divergence coincides with a curvature singularity in Schwarzschild and a caustic in the regular T-duality core. Penrose's theorem~\cite{Penrose:1965} establishes null geodesic incompleteness under its hypotheses; it does not determine the expansion of every congruence. The relevant distinction is whether individual geodesics can be continued, rather than whether the chosen congruence develops an infinite expansion.

\section{Light rays and the case for a photon mass}
\label{sec:null}

\subsection{Radial Maxwell rays}
\label{sec:maxwell}

Consider now an ingoing radial null congruence with conserved energy $E=-k_t$. Its tangent vector is
\begin{equation}
k^{\mu}=\frac{\dd x^{\mu}}{\dd\lambda}=\left(\frac{E}{f},\,-E,\,0,\,0\right),
\label{eq:kmu}
\end{equation}
which is null, $g_{\mu\nu}k^{\mu}k^{\nu}=-E^{2}/f+E^{2}/f=0$, and geodesic. In the language of Eq.~\eqref{eq:Pdef}, $U=0$ and $P=E$: the metric function drops out of the radial equation, $\dd r/\dd\lambda=-E$, and the rays feel no counterpart of the potential that governs the observers. Equations~\eqref{eq:master-theta} and~\eqref{eq:master}, or directly $\theta=(-g)^{-1/2}\partial_\mu(\sqrt{-g}\,k^{\mu})$, give
\begin{equation}
\theta=-\frac{2E}{r},\qquad
\frac{\dd\theta}{\dd\lambda}=-\frac{2E^{2}}{r^{2}} .
\label{eq:null}
\end{equation}
These satisfy the Raychaudhuri equation for a shear-free, twist-free null congruence,
\begin{equation}
\frac{\dd\theta}{\dd\lambda}=-\frac{\theta^{2}}{2}-R_{\mu\nu}k^{\mu}k^{\nu} ,
\label{eq:ray-null}
\end{equation}
because the focusing term vanishes identically: $R^{t}{}_{t}=R^{r}{}_{r}$ implies $R_{\mu\nu}k^{\mu}k^{\nu}=R^{t}{}_{t}\,k_\mu k^{\mu}=0$, or equivalently $R_{\mu\nu}k^{\mu}k^{\nu}=8\pi(\rho+p_r)E^{2}/f=0$ since $p_r=-\rho$. Dividing by $\theta^{2}$ as in Sec.~\ref{sec:tau} gives, exactly and at every radius,
\begin{equation}
\frac{\dd}{\dd\lambda}\!\left(\frac{1}{\theta}\right)=\frac{1}{2}
\quad\Longrightarrow\quad
\theta=-\frac{2}{\lambda_c-\lambda},\qquad \lambda_c=\frac{2}{|\theta_0|} .
\label{eq:null-lambda}
\end{equation}

Equations~\eqref{eq:null} and~\eqref{eq:null-lambda} are identical to those of Minkowski and of Schwarzschild spacetime. The expansion diverges as $r\to0$, which is reached at the finite affine parameter $\lambda_c$, and the expansion alone cannot distinguish the T-duality core from a singular one; what differs is again the nature of the point $r=0$. In Schwarzschild spacetime the rays end there on a curvature singularity and are incomplete. In the T-duality metric the centre is an ordinary point, the divergence is a caustic of the converging congruence, as for a spherical wavefront in flat space, and the rays continue uniquely through the centre (Sec.~\ref{sec:continuation}). On the antipodal side $\theta=+2/(\lambda-\lambda_c)$, the rays cross the inner horizon and the shell into a new exterior region, and $r=E(\lambda-\lambda_c)$ grows without bound, so $\lambda\to\infty$. Radial Maxwell rays are therefore complete. The divergence of $\theta$ is a property of a converging light congruence, not of the spacetime, and requires no remedy.

\subsection{Why a photon mass}
\label{sec:why}

What Eq.~\eqref{eq:null} does show is that radial Maxwell rays are blind to the regular core. The defocusing term $-R_{\mu\nu}u^{\mu}u^{\nu}\to\Leff$, which turns observers back or holds them at the centre, never acts on them, and $D=1/2$ at every radius. Every radial Maxwell ray behaves in the same way: it passes through the centre. The rays form a single class, in contrast to the three classes of observers in Table~\ref{tab:observers}. The reason is that massless rays carry no energy scale. For observers the specific energy $\varepsilon$ is measured in units of the rest mass, and the critical value $\varepsilon=1$ is fixed by $f(0)=1$. For light, $E$ only normalises the affine parameter, $\theta=-2E/r$ for every $E$, and no value of $E$ can single out rays that are screened from the centre or held at it.

A scale of the required kind seems at first to be supplied by the geometry itself. Reference~\cite{Jusufi:2024}, having established timelike completeness for critical observers, noted that for photons near the origin the vector wave operator is modified as
\begin{equation}
\Box \;\longrightarrow\; \nabla_\alpha\nabla^{\alpha} - R^{\nu}{}_{\mu} , \qquad R_{\mu\nu}\simeq\Leff\,g_{\mu\nu} .
\label{eq:JN22}
\end{equation}
Indeed, writing Maxwell's equations $\nabla_\nu F^{\nu\mu}=0$ in terms of the potential and commuting covariant derivatives gives
\begin{equation}
\Box A_\mu - R_\mu{}^{\nu}A_\nu - \nabla_\mu\!\left(\nabla_\nu A^{\nu}\right)=0 ,
\label{eq:maxwell-wave}
\end{equation}
which in Lorenz gauge and in the de Sitter core becomes $(\Box-\Leff)A_\mu=0$. This has the form of a Proca equation with mass squared $\Leff$, and it suggests that the de Sitter core endows light with an effective mass of the order of the inverse de Sitter radius. If that were so, photon rays would become timelike in the core and would inherit the threshold structure of massive observers, with a critical energy set by $\sqrt{\Leff}$. This is the null--timelike correspondence suggested in Ref.~\cite{Jusufi:2024}, and it is the natural starting point for the present work.

The Ricci term is not, however, a physical photon mass, for three reasons. First, it arises from commuting covariant derivatives and is absent from the equivalent first-order equations for $F_{\mu\nu}$, so it is a property of the representation through $A_\mu$ rather than of the field. Second, four-dimensional Maxwell theory is conformally invariant, and the de Sitter core is conformally flat, so Maxwell waves in the core propagate as in flat space, on the light cone. Third, and decisively for rays, the Ricci term does not survive geometric optics. Inserting $A_\mu=a_\mu e^{iS/\epsilon}$ into Eq.~\eqref{eq:maxwell-wave}, the leading order $\epsilon^{-2}$ gives $g^{\mu\nu}\partial_\mu S\,\partial_\nu S=0$, while the Ricci term enters only at order $\epsilon^{0}$. Keeping it at leading order would require local wavenumbers comparable to $\sqrt{\Leff}$, i.e.\ wavelengths of the order of the curvature radius, which is precisely where geometric optics fails. Maxwell rays therefore remain null for every $E$, as Eq.~\eqref{eq:kmu} assumes, and curvature cannot supply the missing scale.

What is required is a term that breaks conformal invariance and that can be retained at the same order as the wavenumber without violating the geometric-optics hierarchy. A genuine dispersive photon mass, generated by a material response, has both properties. In a charged condensate the Anderson--Higgs mechanism yields London screening and a massive vector response~\cite{London:1935,Ginzburg:1950,Weinberg:1986,Anderson:1963,Higgs:1964}. We therefore model the core as a neutral superconducting medium, in which the photon acquires a position-dependent London mass $\mu(r)$. The two terms then play different roles. The Ricci coupling is geometric: it is present for any vector potential on the core, requires no assumption about the source, and acts at the level of the wave equation, but it leaves the light cone unchanged. The London term is material: it requires a charged condensate in the core, is gauge invariant, and breaks conformal invariance. When it is large compared with the curvature, as for $\kappa\gg1$, it can be kept at the same order as the local momentum while curvature corrections remain small, so it survives in the eikonal limit and shapes the rays themselves (Sec.~\ref{sec:waves}). It is therefore the London term, not the Ricci coupling, that realises at the level of rays the null--timelike correspondence suggested in Ref.~\cite{Jusufi:2024}.

With this mass the rays obey Eq.~\eqref{eq:Pdef} with $U=f\mu^{2}$, and the central mass $\mu_0$ sets a critical energy $E=\mu_0$, the counterpart of $\varepsilon=1$ for observers. The complete radial source remains an effective background rather than a microscopic condensate solution. 

\section{London mass in the core}
\label{sec:mass}

To motivate the response and its local core realisation, consider a complex scalar order parameter $\Phi$ of charge $q$ in the relativistic Ginzburg--Landau, or Abelian Higgs, model~\cite{Ginzburg:1950,Higgs:1964}. This action illustrates how the core matter can supply both vacuum energy and an electromagnetic response; it is not an additional material action to be superposed on the T-duality source:
\begin{equation}
S = \int\!\dd^{4}x\,\sqrt{-g}\left[-\frac{F_{\mu\nu}F^{\mu\nu}}{16\pi} - \left|D\Phi\right|^{2} - V(|\Phi|)\right],
\label{eq:action}
\end{equation}
with $D_\mu=\nabla_\mu-i(q/\hbar)A_\mu$. Writing $\Phi=\psi e^{i\vartheta}/\sqrt{2}$, the matter term becomes
\begin{equation}
\left|D\Phi\right|^{2} = \tfrac{1}{2}\,\partial_\mu\psi\,\partial^{\mu}\psi + \frac{q^{2}\psi^{2}}{2\hbar^{2}}\,\cA_\mu\cA^{\mu} ,
\end{equation}
where the phase enters only through the gauge-invariant London momentum
\begin{equation}
\cA_\mu=A_\mu-\frac{\hbar}{q}\,\partial_\mu\vartheta .
\label{eq:calA}
\end{equation}
Varying $A_\mu$ gives $\nabla_\nu F^{\mu\nu}=4\pi J^{\mu}$ with the supercurrent
\begin{equation}
J_\mu = -\frac{q^{2}\psi^{2}}{\hbar^{2}}\,\cA_\mu \equiv -\frac{1}{4\pi\lambda_L^{2}}\,\cA_\mu ,
\label{eq:london}
\end{equation}
which is the covariant London relation~\cite{London:1935,Bhattacharjee:2019,Turimov:2026}; in flat space its spatial and temporal parts reproduce the two London equations. Varying the phase gives charge conservation, $\nabla_\mu(\psi^{2}\cA^{\mu})=0$, and varying the amplitude gives
\begin{equation}
\Box\psi = \frac{q^{2}}{\hbar^{2}}\,\cA_\mu\cA^{\mu}\,\psi + \frac{\dd V}{\dd\psi} .
\label{eq:amplitude}
\end{equation}

In the London limit the amplitude mode is heavy compared with the frequencies of interest, so its perturbation decouples. In the homogeneous central example below the background is $\psi=v$. More generally, a prescribed stiffness profile defines the effective response without assuming that it comes from a canonical scalar profile solving Eq.~\eqref{eq:amplitude}. The linear electromagnetic equations are then 
\begin{equation}
\nabla_\nu F^{\nu\mu} = \mu^{2}(r)\,\cA^{\mu}, \qquad \mu^{2} \equiv \frac{1}{\lambda_L^{2}} = \frac{4\pi q^{2}\psi^{2}}{\hbar^{2}} ,
\label{eq:proca}
\end{equation}
a Proca equation for $\cA_\mu$ with position-dependent mass $m_\gamma=\hbar\mu/c$. Three properties are used below.
\begin{itemize}
\item[(1)] Gauge invariance. No gauge choice removes $\mu^{2}$. In unitary gauge $\cA_\mu=A_\mu$; otherwise the phase acts as a Stueckelberg field.
\item[(2)] Three polarisations. The divergence of Eq.~\eqref{eq:proca} gives
\begin{equation}
\nabla_\mu\!\left(\mu^{2}\cA^{\mu}\right) = 0 \;\;\Longleftrightarrow\;\; \nabla_\mu\cA^{\mu} = -\cA^{\mu}\,\partial_\mu\ln\mu^{2} ,
\label{eq:constraint}
\end{equation}
which removes one component; the condensate phase has become the longitudinal photon.
\item[(3)] Covariant response of the source. The London law contains a scalar stiffness $\mu^2$ and no explicit medium four-velocity, so its equations remain defined between the horizons. We interpret it as a particular relativistic response of the existing source, rather than as the general response tensor of an ordinary particle plasma. The complete microscopic source is not constructed here.
\end{itemize}

For the radial propagation problem we use the following quadratic response action about the zero-field, zero-current core state. In this action and the subsequent wave and ray equations, $A_\mu$ and $\cA_\mu$ denote electromagnetic perturbations and their gauge-invariant combination:
\begin{equation}
S_{\rm resp}^{(2)} = \int\!\dd^{4}x\,\sqrt{-g}\left[-\frac{F_{\mu\nu}F^{\mu\nu}}{16\pi} - \frac{\mu^{2}(r)}{8\pi}\,\cA_\mu\cA^{\mu}\right],
\label{eq:Sresp}
\end{equation}
Here $\mu^{2}(r)$ is the prescribed stiffness evaluated on the background source. Variation with respect to the electromagnetic perturbation and phase gives Eqs.~\eqref{eq:proca} and~\eqref{eq:constraint}. The metric already solves $G_{\mu\nu}=8\pi T^{\rm (core)}_{\mu\nu}$ with the density and pressures of Eqs.~\eqref{eq:rho} and~\eqref{eq:pperp}. The response action adds electromagnetic fluctuations about that source; it does not add another copy of its background energy. As in effective plasma-ray descriptions~\cite{Perlick:2017,Perlick:2022}, the propagation coefficient is prescribed. Unlike a separate plasma cloud, the responding material here is identified with the source itself. Section~\ref{sec:backreaction} establishes the background counting, gives a central potential construction, and states its limitations.

As a constitutive assumption we take the coefficient to follow the T-duality source,
\begin{equation}
\mu^{2}(r) = 4\pi\kappa\,\rho(r) = \frac{3\kappa M \lz^{2}}{\left(r^{2}+\lz^{2}\right)^{5/2}},
\label{eq:mu}
\end{equation}
with $\kappa>0$ the dimensionless response parameter of the propagation model. The local potential example below also contains an amplitude self-coupling, which does not enter the ray equations. Since $p_r=-\rho$, $\rho$ is invariant under radial boosts, so Eq.~\eqref{eq:mu} defines a scalar across both horizons, as property (3) requires. All results below are conditional on the prescribed-response model~\eqref{eq:Sresp} with~\eqref{eq:mu}. They do not derive a superconducting transition or a microscopic action producing the complete radial background and response. The central potential construction supplies a local consistency example only.

The mass is generated by the medium; curvature enters only through Eq.~\eqref{eq:mu}. With $G^{t}{}_{t}=-8\pi\rho$ it becomes a statement about geometry,
\begin{equation}
\mu^{2}(r) = -\frac{\kappa}{2}\,G^{t}{}_{t}(r) ,
\label{eq:muG}
\end{equation}
Equation~\eqref{eq:muG} is an identity on the stipulated background, not a prescription to replace the source stiffness by an Einstein-tensor component before varying an action. For $\kappa=\mathcal{O}(1)$ the photon mass is of the order of the local curvature scale. At the centre,
\begin{equation}
\mu_0^{2} = \frac{3\kappa M}{\lz^{3}} = \frac{\kappa}{2}\,\Leff ,
\label{eq:mu0}
\end{equation}
and therefore
\begin{equation}
\frac{\lambda_L(0)}{\LdS} = \sqrt{\frac{2}{3\kappa}} .
\label{eq:ratio}
\end{equation}
This ratio depends neither on $M$ nor on $\lz$, and it is not special to the T-duality profile. For any regular core with central density $\rho_0$ one has $\LdS^{2}=3/(8\pi\rho_0)$ and $\mu_0^{2}=4\pi\kappa\rho_0$, hence $\mu_0^{2}\LdS^{2}=3\kappa/2$. Thus, within the constitutive family $\mu^2=4\pi\kappa\rho$ at fixed $\kappa$, the central penetration depth is comparable to or shorter than the de Sitter radius for $\kappa\gtrsim1$. This conditional universality does not establish that an arbitrary regular core condenses or obeys the same constitutive law.

The screening picture of Ref.~\cite{Jusufi:2024} has an electromagnetic counterpart. There the vacuum screens gravity over $\lz$, as charges screen an electrostatic field over the Debye length. Here the same density profile screens electromagnetism over the London depth, with
\begin{equation}
\frac{\lambda_L(0)}{\lz} = \sqrt{\frac{\lz}{3\kappa M}} ,
\label{eq:screen}
\end{equation}
so for $\kappa M\gg\lz$ electromagnetic screening acts on shorter scales than gravitational screening.

\section{Core response and central potential}
\label{sec:backreaction}

\subsection{One background source and its electromagnetic response}

The density $\rho$ in Eq.~\eqref{eq:rho} is the energy density of the material already sourcing the geometry. We assume that this material itself possesses the response~\eqref{eq:mu}. Its potential, gradient and interaction stresses, if present in a microscopic description, must be constituents of $T^{\rm (core)}_{\mu\nu}$ rather than a second stress tensor added on top of it. At zero electromagnetic perturbation,
\begin{equation}
T^{\rm (background)}_{\mu\nu}=T^{\rm (core)}_{\mu\nu},
\qquad G_{\mu\nu}=8\pi T^{\rm (core)}_{\mu\nu}.
\label{eq:single-source}
\end{equation}
No separate dilute carrier component is introduced, and no extra density is added to $\rho$.

The neutral background obeys $\bar F_{\mu\nu}=0$ and $\bar{\cA}_\mu=0$, hence $\bar J_\mu=0$, but its stiffness can be nonzero. A charged order parameter need not represent a gas with nonzero net charge density: the current is proportional to the gauge-invariant phase momentum, whereas the London coefficient is proportional to the squared amplitude. Thus a zero-current charged-field condensate can produce a massive electromagnetic response without a background electric field or an additional neutralising fluid. A finite-density microscopic realisation would require a separate analysis; it is not assumed here.

Because $S_{\rm resp}^{(2)}$ is quadratic in the electromagnetic perturbations,
\begin{equation}
S_{\rm resp}^{(2)}\big|_{A=0,\,\cA=0}=0,
\qquad T^{\rm (resp)}_{\mu\nu}\big|_{A=0,\,\cA=0}=0.
\label{eq:zero-response-background}
\end{equation}
Any variation of the source-dependent coefficient multiplies a quadratic perturbation and therefore also vanishes on this background. Electromagnetic perturbations thus contribute stress only at quadratic order in their amplitude and propagate as test fields on the fixed geometry. This statement assumes expansion about a background solution of the underlying source theory; it does not supply that theory. Vacuum loop corrections are not computed and, in an effective description, require renormalised background parameters.

\subsection{A potential reproducing the leading de Sitter core}

A simple local example uses the canonical charged scalar of Eq.~\eqref{eq:action}, with $V$ expressed in terms of the real amplitude $\psi$:
\begin{equation}
V(\psi)=\rho_0+\frac{\lambda_\psi}{4}(\psi^2-v^2)^2,
\qquad \rho_0=\frac{3M}{4\pi\lz^3},
\label{eq:core-potential}
\end{equation}
where $\lambda_\psi>0$ and $v>0$. The homogeneous configuration
\begin{equation}
\psi=v,\qquad \cA_\mu=0,\qquad F_{\mu\nu}=0
\label{eq:core-condensate}
\end{equation}
solves the matter equations, since $V'(v)=0$. Its stress tensor and central curvature are
\begin{equation}
T^{(\psi)}_{\mu\nu}=-\rho_0 g_{\mu\nu},
\qquad \Leff=8\pi\rho_0=\frac{6M}{\lz^3}.
\label{eq:potential-core-stress}
\end{equation}
Identifying this vacuum energy with the existing central source density, rather than adding it to that density, reproduces the leading de Sitter core. The same state gives
\begin{equation}
\mu_0^2=\frac{4\pi q^2v^2}{\hbar^2},
\qquad v^2=\frac{\kappa\hbar^2\rho_0}{q^2},
\label{eq:core-vev}
\end{equation}
so $\mu_0^2=4\pi\kappa\rho_0=\kappa\Leff/2$, as required. The amplitude fluctuation has inverse Compton length $\mathfrak m_H$ given by
\begin{equation}
\mathfrak m_H^2=V''(v)=2\lambda_\psi v^2.
\label{eq:higgs-scale}
\end{equation}
The London approximation requires this scale to exceed the local photon frequencies and the inverse background variation lengths in the region of application.

This is an exact reconstruction of the source at the centre, where the T-duality stress is isotropic vacuum energy, but not of the full T-duality solution. In particular, a constant $v$ gives constant $\mu_0$, and does not derive the $O(r^2)$ variation of $\mu^2$ or its density-gradient force. Those terms in the subsequent ray analysis follow from the separately prescribed full profile~\eqref{eq:mu}. The parameters in Eq.~\eqref{eq:core-potential} match one chosen central background; a single microscopic potential generating the entire family parametrised by $M$ has not been demonstrated. The symmetry-breaking scale and self-coupling are local illustrative parameters, not predictions of T-duality.

\subsection{Why the potential alone does not reconstruct the full source}

The energy density $\rho_\psi$ and pressures of the condensate are computed from its own action and are not, in general, those of the T-duality source. If the condensate were the entire source, they would have to coincide with it at every radius, $\rho_\psi=\rho$, $p_{r,\psi}=p_r=-\rho$ and $p_{\perp,\psi}=p_\perp$. To test this, consider a static canonical scalar in a region with $f>0$ and define $K_r=f\psi'^2/2$ and $K_t=q^2\psi^2\cA_t^2/(2\hbar^2 f)$. Its stress components are
\begin{align}
\rho_\psi&=K_r+K_t+V,\nonumber\\
p_{r,\psi}&=K_r+K_t-V,\nonumber\\
p_{\perp,\psi}&=-K_r+K_t-V.
\label{eq:scalar-stresses}
\end{align}
Hence
\begin{equation}
\rho_\psi+p_{r,\psi}
=f\psi'^2+\frac{q^2\psi^2\cA_t^2}{\hbar^2 f}\geq0.
\label{eq:NECpsi}
\end{equation}
The potential cancels from this identity. For a nonzero condensate to satisfy the required $p_r=-\rho$, both $\psi'=0$ and $\cA_t=0$ must hold in a static region. The resulting stress is isotropic vacuum energy, whereas the full source obeys
\begin{equation}
\rho+p_\perp=\frac{5r^2\rho}{2(r^2+\lz^2)}>0
\quad (r>0).
\label{eq:source-anisotropy}
\end{equation}
Thus the canonical potential construction matches the leading central stress but cannot reproduce the entire radial source. The mismatch is controlled by the anisotropy~\eqref{eq:source-anisotropy}, which grows only slowly away from the centre,
\begin{equation}
\frac{\rho+p_\perp}{\rho}=\frac{5r^{2}}{2\left(r^{2}+\lz^{2}\right)}\simeq\frac{5}{2}\,\frac{r^{2}}{\lz^{2}}\qquad (r\ll\lz) .
\label{eq:anisotropy-small-r}
\end{equation}
The identification of the core with a superconducting condensate is therefore exact at $r=0$, where $\rho_\psi=V(v)=\rho_0=\rho(0)$, and approximate throughout the inner de Sitter region $r\ll\lz$, with corrections of relative order $r^{2}/\lz^{2}$. It fails for $r\sim\lz$ and beyond, where the source is strongly anisotropic. The obstruction is not specific to the canonical action. For any theory of scalar fields with Lagrangian $L(X,\psi)$, a static configuration depending only on $r$ has $T^{\mu}{}_{\nu}=L_X\,\partial^{\mu}\psi\,\partial_{\nu}\psi-\delta^{\mu}{}_{\nu}L$, whose time and angular components carry no gradient term. Hence $T^{t}{}_{t}=T^{\vartheta}{}_{\vartheta}$, i.e.\ $p_\perp=-\rho$ identically, which is incompatible with Eq.~\eqref{eq:source-anisotropy} for any choice of potential or kinetic function. Related no-go results for minimally coupled real scalar sources also show why an arbitrary potential does not guarantee a regular black hole~\cite{Bronnikov:2001}. A microscopic completion would require a more general source sector, for example additional anisotropic fields, nonminimal interactions or nonlocal dynamics, with all of its field equations and stresses solved together. No such completion is claimed here. Away from the inner core, identifying the source as superconducting is an effective constitutive assumption, not a solution of any static scalar theory.

\section{Photon propagation in the medium}
\label{sec:waves}

The London depth grows outward,
\begin{equation}
\lambda_L(r) = \lambda_L(0)\left(1+\frac{r^{2}}{\lz^{2}}\right)^{5/4}, \qquad \lambda_L(0) = \sqrt{\frac{\lz^{3}}{3\kappa M}} ,
\end{equation}
so the medium is effectively confined to $r\lesssim\lz$. At the outer horizon of a large black hole $\mu^{2}(r_+)\simeq3\kappa\lz^{2}/(32M^{4})$, smaller than $\mu_0^{2}$ by a factor $32(M/\lz)^{5}$. In ordinary units, with $\lz$ equal to the Planck length, the central and horizon values are
\begin{align}
m_\gamma(0) &= \sqrt{3\kappa\,\frac{M}{\MPl}}\;\MPl , \nonumber\\
m_\gamma(r_+) &\simeq \sqrt{\frac{3\kappa}{32}}\left(\frac{\MPl}{M}\right)^{2}\MPl .
\label{eq:units}
\end{align}
For a solar-mass black hole the horizon value is $\sim5\times10^{-49}\sqrt{\kappa}$~eV, far below experimental bounds, whereas the central value is $\sim2\times10^{19}\sqrt{\kappa}\,\MPl$. The latter is trans-Planckian because the core curvature is: $\LdS<\lz$ whenever $\lz<2M$. Near the extremal mass all scales are comparable, $\mu_0\lz\simeq2\sqrt{\kappa}$.

For a static observer, who exists in the core and in the exterior, a wave of Killing frequency $E$ has local frequency $\omega_{\rm loc}=E/\sqrt{f}$ and obeys
\begin{equation}
\omega_{\rm loc}^{2} = k^{2} + \mu^{2}(r) .
\label{eq:disp}
\end{equation}
Its group velocity is
\begin{equation}
v_g = \frac{k}{\omega_{\rm loc}} = \sqrt{1-\frac{f\mu^{2}}{E^{2}}} .
\label{eq:vg}
\end{equation}
This local transverse dispersion relation also occurs in a collisionless plasma with $\varepsilon=1-\mu^2/\omega_{\rm loc}^2$; it does not identify the longitudinal Proca mode with an ordinary plasma mode. Where a local homogeneous approximation is valid, $E^2<f\mu^2$ gives an imaginary local radial wavenumber. Since $f\mu^2\leq\mu_0^2$, with equality only at $r=0$, and $f\mu^2<0$ between the horizons, $\mu_0$ is the threshold for a radial Hamiltonian ray to reach the centre. Physical spherical waves also contain the mode-dependent terms derived below.

Static fields are screened. For a static axial multipole $\cA_\varphi\propto\Psi(r)\sin\theta\,\partial_\theta P_\ell(\cos\theta)$ in the de Sitter core, Eq.~\eqref{eq:proca} reduces to $(f\Psi')'=[\ell(\ell+1)/r^{2}+\mu_0^{2}]\Psi$. The solution regular at the centre is
\begin{equation}
\Psi = r^{\ell+1}\,{}_2F_1\!\left(a_+,a_-;\ell+\tfrac{3}{2};\frac{r^{2}}{\LdS^{2}}\right),
\label{eq:static}
\end{equation}
with
\begin{equation}
a_\pm = \frac{1}{2}\left[\ell+\frac{3}{2}\pm\sqrt{\frac{1}{4}-\frac{3\kappa}{2}}\,\right] .
\end{equation}
For $\kappa\gg1$ one has $\lambda_L\ll\LdS$, and the solution tends to the flat-space Meissner profile $\Psi\propto r\,i_\ell(\mu_0 r)$: the field at the centre is suppressed relative to its value at radius $r$ by $e^{-\mu_0 r}$, up to powers of $r$. Since the third argument of ${}_2F_1$ equals $a_++a_-$, $\Psi$ diverges logarithmically at $r=\LdS$; this happens already at $\kappa=0$ and reflects the familiar absence of static hair in a de Sitter patch.

The prescribed response profile alone does not fix a superconducting critical field. In the homogeneous potential example~\eqref{eq:core-potential}, the free-energy difference between the uncondensed configuration $\psi=0$ and $\psi=v$ is
\begin{equation}
\Delta V=V(0)-V(v)=\frac{\lambda_\psi v^4}{4}.
\label{eq:condensation-energy}
\end{equation}
It is distinct from the vacuum-energy offset $\rho_0$. In a local homogeneous thermodynamic approximation with the usual magnetic free-energy balance, one would obtain $B_c^2/(8\pi)=\Delta V$, or $B_c^2=2\pi\lambda_\psi v^4$. This estimate is conditional on that approximation and does not derive a radial critical-field profile, interface physics, vortex thresholds or the stability of the full source. The independent test-field requirement is that the energy density of the electromagnetic perturbation remain small compared with $\rho$.

Treating $\cA_\mu$ as a test field and commuting derivatives in Eq.~\eqref{eq:proca}, we find
\begin{equation}
\Box\cA_\mu - R_\mu{}^{\nu}\cA_\nu - \mu^{2}\cA_\mu = \nabla_\mu\!\left(\nabla_\nu\cA^{\nu}\right) ,
\label{eq:wave}
\end{equation}
with the right-hand side fixed by Eq.~\eqref{eq:constraint}; it vanishes where $\mu$ is uniform. Axial perturbations are divergence-free on the sphere, so no derivative of $\mu$ enters. With $\dd r_*=\dd r/f$ and time dependence $e^{-i\omega t}$ they obey
\begin{equation}
\frac{\dd^{2}\Psi}{\dd r_*^{2}} + \left[\omega^{2} - V_{\rm ax}\right]\Psi = 0,
\label{eq:axial}
\end{equation}
with potential
\begin{equation}
V_{\rm ax} = f\left[\frac{\ell(\ell+1)}{r^{2}} + \mu^{2}(r)\right] ,
\label{eq:Vax}
\end{equation}
which reduces to that of Maxwell fields on the T-duality black hole at $\kappa=0$.

For polar perturbations, $\cA_\mu\dd x^\mu=[b_0\dd t+b_1\dd r]Y_{\ell m}+b_2\partial_aY_{\ell m}\dd x^{a}$, the constraint~\eqref{eq:constraint} reads
\begin{equation}
-\frac{\mu^{2}}{f}\,\dot b_0 + \frac{1}{r^{2}}\left(r^{2} f \mu^{2} b_1\right)' - \frac{\ell(\ell+1)\mu^{2}}{r^{2}}\,b_2 = 0 .
\end{equation}
It differs from the constant-mass case~\cite{Rosa:2012} by terms in $\mu'/\mu$ that couple the transverse and longitudinal modes to the density gradient. For $\ell=0$ only the longitudinal mode survives: the radial electric field $\mathcal{E}=\dot b_1-b_0'=\mu\Phi/r$ obeys Eq.~\eqref{eq:axial} with
\begin{equation}
V_{0}(r) = f\mu^{2} + \mu r\,\frac{\dd^{2}}{\dd r_*^{2}}\!\left(\frac{1}{\mu r}\right) .
\end{equation}
This longitudinal monopole mode of the massive photon has no massless analogue. It is not an oscillation of the condensate amplitude, which is frozen in the London limit.

Inserting $\cA_\mu=a_\mu e^{iS}$, with $p_\mu=\partial_\mu S$ and $\mu$ counted on the same footing as $p_\mu$, the leading eikonal order gives
\begin{equation}
g^{\mu\nu}p_\mu p_\nu + \mu^{2}(r) = 0 ,
\label{eq:eikonal}
\end{equation}
the covariant form of Eq.~\eqref{eq:disp}. This is a massive-particle WKB scaling: the local momentum and $\mu$ are retained at the same order, while curvature, coefficient derivatives and amplitude derivatives are smaller. In addition to slow variation of the local wavenumber, one needs curvature components small compared with $\omega_{\rm loc}^2$. Near $E\sim\mu_0$ in the core, a sufficient curvature hierarchy requires $\kappa\gg1$, together with control of mass-gradient corrections. In the homogeneous example of Sec.~\ref{sec:backreaction}, the amplitude mode must also be heavy enough to justify the London limit, Eq.~\eqref{eq:higgs-scale}. The Hamiltonian ray results themselves hold for any $\kappa>0$; the condition $\kappa\gg1$ concerns their interpretation as localised wave packets. In the strict high-frequency limit at fixed $\mu$, the principal characteristics of the Proca PDE are null; the timelike trajectories here instead describe dispersive wave packets in the stated semiclassical scaling.

In the core the London and Ricci terms appear together. Equation~\eqref{eq:wave} contains the Ricci coupling of Eq.~\eqref{eq:JN22}, and at $\kappa=0$ and in Lorenz gauge it reduces to the Maxwell equation $(\Box-\Leff)A_\mu=0$. With the medium present and in a homogeneous-medium approximation, $\mu\simeq\mu_0$, the core equation becomes
\begin{equation}
\left[\Box - \Leff - \mu_0^{2}\right]\cA_\mu = 0 , \qquad \nabla_\mu\cA^{\mu}=0 ,
\label{eq:coreeq}
\end{equation}
and the coefficient of the non-derivative term is $\Leff(1+\kappa/2)$. The approximation drops the right-hand side of Eq.~\eqref{eq:wave}, which does not vanish even at the centre. There $\partial_\mu\mu^{2}=0$ but $\partial_i\partial_j\ln\mu^{2}|_0=-5\delta_{ij}/\lz^{2}$ in regular local coordinates, and the constraint~\eqref{eq:constraint} gives
\begin{equation}
\nabla_i\!\left(\nabla_\nu\cA^{\nu}\right)\big|_0 = \frac{5}{\lz^{2}}\,\cA_i\big|_0 ,
\label{eq:gradcorr}
\end{equation}
an additional coefficient for the spatial components. Relative to the London term it is $5\lz/(3\kappa M)$, so Eq.~\eqref{eq:coreeq} holds for $\kappa M\gg\lz$. The sum $\Leff+\mu_0^{2}$ is a coefficient of the operator, not a total photon mass squared; the physical mass parameter of the photon is $\mu$ alone, and only $\mu$ survives in the eikonal limit~\eqref{eq:eikonal}, as argued in Sec.~\ref{sec:why}. The two coefficients are equal when
\begin{equation}
\mu_0^{2} = \Leff \quad\Longleftrightarrow\quad \kappa = 2 ,
\label{eq:kappa2}
\end{equation}
and by Eq.~\eqref{eq:muG} the London coefficient then tracks the curvature at every radius, not only at the centre. We note this coincidence without attributing to it any established physical significance.

\section{Massive-photon rays}
\label{sec:rays}

Equation~\eqref{eq:eikonal} is the constraint $\mathcal{H}=0$ of the Hamiltonian~\cite{Synge:1960,Perlick:2017}
\begin{equation}
\mathcal{H} = \tfrac{1}{2}\left[g^{\mu\nu}p_\mu p_\nu + \mu^{2}(r)\right] .
\label{eq:H}
\end{equation}
With $\dd x^{\mu}/\dd\lambda=p^{\mu}$, Hamilton's equations give
\begin{equation}
p^{\nu}\nabla_\nu p_\mu = -\tfrac{1}{2}\,\partial_\mu\mu^{2} .
\label{eq:force}
\end{equation}
The rays are timelike, $p^{2}=-\mu^{2}$, and feel a force down the gradient of the plasma density. Since $\mu^{2}$ decreases outward, the medium pushes photons away from the centre, like the de Sitter repulsion.

Equivalently, the rays extremise $-\int\mu\,\dd s$, with $\dd s$ the proper time of $g_{\mu\nu}$, and are, up to reparametrisation, timelike geodesics of the conformal metric
\begin{equation}
\tilde g_{\mu\nu} = \mu^{2}(r)\,g_{\mu\nu} .
\label{eq:gtilde}
\end{equation}
Two parameters along the ray must be distinguished from $\lambda$. The proper time of the physical metric is $\dd\tau=\mu\,\dd\lambda$, since $-g_{\mu\nu}p^{\mu}p^{\nu}=\mu^{2}$. The affine parameter of $\tilde g_{\mu\nu}$, its proper time, is instead
\begin{equation}
\dd\tilde\tau = \mu^{2}\,\dd\lambda ,
\label{eq:tautilde}
\end{equation}
because the unit-mass Hamiltonian of $\tilde g_{\mu\nu}$ is $\tilde{\mathcal{H}}=\tfrac12(\tilde g^{\mu\nu}p_\mu p_\nu+1)=\mathcal{H}/\mu^{2}$, whose flow on $\mathcal{H}=0$ is that of $\mathcal{H}$ slowed down by $\mu^{2}$. The conformal factor is smooth, even in $r$ and strictly positive at finite radius, so the map onto $\tilde g_{\mu\nu}$ fixes the shape of the rays; this is the ray-level form of the correspondence discussed in Sec.~\ref{sec:why}. Completeness, however, is not a statement about $\tilde g_{\mu\nu}$: because $\mu^{2}\to0$ at large $r$, $\tilde g_{\mu\nu}$ is itself timelike-incomplete at infinity (Sec.~\ref{sec:complete}), and photon completeness must be judged with $\lambda$.

With the conserved energy $E=-p_t$ and angular momentum $J=p_\varphi$, the radial equation is
\begin{equation}
\left(\frac{\dd r}{\dd\lambda}\right)^{2} = E^{2} - f\left[\frac{J^{2}}{r^{2}} + \mu^{2}\right] .
\label{eq:radial}
\end{equation}
The effective potential is $V_{\rm ax}$ with $\ell(\ell+1)\to J^{2}$; for radial rays it reduces to $U=f\mu^{2}$. Radial timelike geodesics obey the same equation with $U=f$, so the photon problem follows from that of Sec.~\ref{sec:background} by the replacement $f\to f\mu^{2}$ and $\varepsilon\to E$.

For $J\neq0$ the centrifugal barrier keeps rays away from the centre. For radial rays the potential near the centre is
\begin{equation}
f\mu^{2} \simeq \mu_0^{2} - \sigma^{2}r^{2} ,
\label{eq:Uexp}
\end{equation}
with
\begin{equation}
\sigma^{2} \equiv \mu_0^{2}\left(\frac{1}{\LdS^{2}} + \frac{5}{2\lz^{2}}\right) .
\label{eq:sigma}
\end{equation}
The first term is the de Sitter repulsion also felt by massive observers; the second is the plasma-gradient force of Eq.~\eqref{eq:force}. Three cases arise.

(1) Supercritical, $E>\mu_0$. The ray crosses $r=0$ with $\dd r/\dd\lambda=-\sqrt{E^{2}-\mu_0^{2}}$, at finite $\lambda$ and $\tau$, and continues antipodally, with group velocity $v_g=\sqrt{1-\mu_0^{2}/E^{2}}$ at the centre.

(2) Subcritical, $E<\mu_0$. The ray is reflected where $f\mu^{2}=E^{2}$, at
\begin{equation}
r_t \simeq \frac{\mu_0}{\sigma}\sqrt{1 - \frac{E^{2}}{\mu_0^{2}}} \;\xrightarrow{\;M\gg\lz\;}\; \LdS\sqrt{1 - \frac{E^{2}}{\mu_0^{2}}} ,
\label{eq:rt}
\end{equation}
returns to $r_-$ and leaves through the future inner horizon, as subcritical observers do. The radial Hamiltonian trajectory does not enter $r<r_t$. The wave solution must instead be obtained with its mode-dependent potential and regular central boundary condition; the ray exclusion is not a proof of exact wave shielding.

(3) Critical, $E=\mu_0$. The centre is an unstable equilibrium at which a photon can sit at rest, $p^{\mu}=(\mu_0,0,0,0)$. A ray with this energy approaches it as
\begin{equation}
r\propto e^{-\sigma\lambda}=e^{-(\sigma/\mu_0)\tau},
\end{equation}
in infinite $\lambda$, and hence also in infinite $\tau$ and $\tilde\tau$.

The three cases are the photon counterparts of the three classes of observers in Sec.~\ref{sec:classes}; in particular, case (3) corresponds to the critical observers of Ref.~\cite{Jusufi:2024}, who approach the centre as $e^{-\tau/\LdS}$. For photons the rate $1/\LdS$ is replaced by $\sigma/\mu_0=[\LdS^{-2}+5/(2\lz^{2})]^{1/2}$, independent of $\kappa$. At $\kappa=0$ the shielding of case (2) disappears and every radial ray crosses the centre.

The quantity $k_*^2=E^2-f\mu^2$ is the radial ray momentum squared, rather than the exact wavenumber squared of a spherical Proca mode. In particular, axial modes have $\ell\geq1$ and retain the angular term in Eq.~\eqref{eq:Vax}. For the physical longitudinal monopole, expanding its exact potential gives
\begin{equation}
V_0(r)=\frac{2}{r^2}+\mu_0^2-\frac{2}{\LdS^2}-\frac{5}{2\lz^2}+O(r^2).
\label{eq:V0centre}
\end{equation}
The $2/r^2$ term is a feature of the radial master variable, not a curvature singularity. The regular branch has $\Phi\propto r^2$, so $\mathcal E=\mu\Phi/r=O(r)$ at the centre; the singular branch $\Phi\propto r^{-1}$ is excluded. The physical near-centre mode equation is therefore not a parabolic-barrier equation with potential $f\mu^2$ alone. We infer neither a universal full-wave transmission threshold at $E=\mu_0$ nor a universal threshold width from the ray potential. The crossing, reflection and asymptotic approach described above are exact properties of the prescribed Hamiltonian rays. Their relation to a finite wave packet near the endpoint requires solving the appropriate Proca mode equation with regularity and horizon boundary conditions.

\section{Raychaudhuri equation and completeness for massive photons}
\label{sec:complete}

We now repeat the analysis of Sec.~\ref{sec:background} for the massive photons of Sec.~\ref{sec:rays}. Consider a congruence $p_\mu=\partial_\mu S$ and define $\theta=\nabla_\mu p^{\mu}$ and $B_{\mu\nu}=\nabla_\mu p_\nu$. Here $\theta$ is the divergence of the unnormalised momentum, not the proper-time expansion of $u^{\mu}=p^{\mu}/\mu$; the two are related by $\Theta_u=(\theta-p^{\mu}\partial_\mu\ln\mu)/\mu$ and coincide, up to the factor $\mu_0$, at the centre, where $\partial_\mu\mu=0$. Differentiating along the rays and using Eq.~\eqref{eq:force},
\begin{equation}
\frac{\dd\theta}{\dd\lambda} = -B_{\mu\nu}B^{\mu\nu} - R_{\mu\nu}\,p^{\mu}p^{\nu} - \tfrac{1}{2}\,\Box\mu^{2} .
\label{eq:ray-photon}
\end{equation}
Compared with Eq.~\eqref{eq:ray-JN} there is one new term, the d'Alembertian of the plasma density, which originates in the force of Eq.~\eqref{eq:force}. The curvature term is tied to that of the observers: since $R^{t}{}_{t}=R^{r}{}_{r}$, any radial vector with $p^{2}=-\mu^{2}$ gives, for every $E$,
\begin{equation}
R_{\mu\nu}\,p^{\mu}p^{\nu} = \mu^{2}\,R_{\mu\nu}u^{\mu}u^{\nu} = 8\pi\mu^{2}\pperp ,
\label{eq:Rpp}
\end{equation}
which changes sign at the same radius, $r=\sqrt{2/3}\,\lz$, where the strong energy condition fails.

For radial rays, $p^{r}=\dd r/\dd\lambda$ obeys Eq.~\eqref{eq:Pdef} with $s=\lambda$, $\mathcal{E}=E$ and $U=f\mu^{2}$, so $\theta$ and $\dd\theta/\dd\lambda$ are given by Eqs.~\eqref{eq:master-theta} and~\eqref{eq:master} with $P_E=\sqrt{E^{2}-f\mu^{2}}$. Dividing Eq.~\eqref{eq:ray-photon} by $\theta^{2}$ gives the photon counterpart of Eq.~\eqref{eq:inv-theta},
\begin{equation}
\frac{\dd}{\dd\lambda}\!\left(\frac{1}{\theta}\right)=\frac{B_{\mu\nu}B^{\mu\nu}+R_{\mu\nu}p^{\mu}p^{\nu}+\tfrac12\Box\mu^{2}}{\theta^{2}}\equiv D_E ,
\label{eq:inv-theta-photon}
\end{equation}
and, as in Eq.~\eqref{eq:tau-int},
\begin{equation}
\lambda=\int_{1/\theta_0}^{1/\theta}\frac{\dd\left(\theta^{-1}\right)}{D_E} .
\label{eq:lambda-int}
\end{equation}
Near the centre the potential is $f\mu^{2}\simeq\mu_0^{2}-\sigma^{2}r^{2}$ [Eq.~\eqref{eq:Uexp}], so $P_E^{2}\simeq E^{2}-\mu_0^{2}+\sigma^{2}r^{2}$. This has the same form as $P_\varepsilon^{2}$ for observers, with $\varepsilon^{2}-1\to E^{2}-\mu_0^{2}$ and $1/\LdS^{2}\to\sigma^{2}$, and the analysis of Sec.~\ref{sec:classes} carries over.

\subsection{Critical photons}
\label{sec:crit-photon}

For $E=\mu_0$, Eqs.~\eqref{eq:master-theta} and~\eqref{eq:master} give, near the centre,
\begin{equation}
\theta \simeq -3\sigma, \qquad
\frac{\dd\theta}{\dd\lambda} \simeq -\frac{5\mu_0^{2}\left(64M+35\,\lz\right)}{8\,\lz^{5}}\;r^{2} .
\label{eq:theta0}
\end{equation}
The expansion is finite and its rate vanishes quadratically, as for the critical observers, Eqs.~\eqref{eq:Theta-JN} and~\eqref{eq:dTheta-JN}, and $D_{\mu_0}\to0$. Since $\theta-\theta(0)\propto r^{2}$ and $\dd r/\dd\lambda\simeq-\sigma r$, Eq.~\eqref{eq:ray-photon} linearises to $\dd\theta/\dd\lambda\simeq-2\sigma\left[\theta-\theta(0)\right]$, and
\begin{equation}
\lambda-\lambda_1 \simeq -\frac{1}{2\sigma}\,\ln\frac{\theta-\theta(0)}{\theta_1-\theta(0)} = \frac{1}{\sigma}\,\ln\frac{r_1}{r} \longrightarrow \infty
\label{eq:lambda-crit}
\end{equation}
as $r\to0$. The critical rays reach the centre only at infinite $\lambda$, as in Eq.~\eqref{eq:tau-JN} with $1/\LdS$ replaced by $\sigma$. For $M=5\,\lz$ the rate is most negative at $r\simeq0.45\,\lz$ and returns to zero at the centre, the behaviour that Ref.~\cite{Jusufi:2024} associates with the onset of repulsive gravity.

The vanishing of the rate is a cancellation among three terms:
\begin{align}
-R_{\mu\nu}\,p^{\mu}p^{\nu} & \rightsquigarrow \Leff\,\mu_0^{2},\\
-\tfrac{1}{2}\,\Box\mu^{2} & \rightsquigarrow \frac{15}{2\lz^{2}}\,\mu_0^{2},\\
-B_{\mu\nu}B^{\mu\nu} & \rightsquigarrow -\left(\Leff + \frac{15}{2\lz^{2}}\right)\mu_0^{2} .
\label{eq:cancel}
\end{align}
The third limit follows from the radial deformation tensor. With $p_\mu=(-E,-P_E/f,0,0)$, its contraction away from the static-coordinate horizons is
\begin{equation}
B_{\alpha\beta}B^{\alpha\beta}
=P_E'^2-\frac{f'}{f}P_EP_E'
-\frac{\mu^2 f'^2}{2f}+\frac{2P_E^2}{r^2}.
\label{eq:Bcontract}
\end{equation}
For the critical family, $P_E=\sigma r+O(r^3)$ and $f'=O(r)$; the first and last terms tend to $\sigma^2$ and $2\sigma^2$, while the others vanish, so $B_{\alpha\beta}B^{\alpha\beta}\to3\sigma^2=\mu_0^2[\Leff+15/(2\lz^2)]$. The first line of Eq.~\eqref{eq:cancel} is the de Sitter defocusing of Ref.~\cite{Jusufi:2024}; the second is a defocusing by the medium, positive because the plasma density peaks at the centre. It is local: at $r=\lz$ one finds $-\tfrac12\Box\mu^{2}=-5\mu^{2}/(8\lz^{2})<0$, so the medium term focuses there. For $M\gg\lz$ the curvature term dominates by a factor $4M/5\lz$, while near the extremal mass the two are comparable.

\subsection{Supercritical and subcritical photons}
\label{sec:super-photon}

For $E>\mu_0$, $P_E\to P_0=\sqrt{E^{2}-\mu_0^{2}}>0$ and
\begin{equation}
\theta=-\frac{2P_0}{r}-\frac{2\sigma^{2}r}{P_0}+O(r^3), \qquad D_E\to\frac{1}{2} .
\label{eq:theta-super-photon}
\end{equation}
Integrating Eq.~\eqref{eq:lambda-int}, the affine parameter from $\theta_0$ to the centre is
\begin{equation}
\lambda_c=\frac{2}{|\theta_0|}+\frac{20\,\sigma^{2}}{3|\theta_0|^{3}}+O\!\left(|\theta_0|^{-5}\right) ,
\label{eq:lambda-caustic}
\end{equation}
the counterpart of Eq.~\eqref{eq:caustic-tau}. The rays reach the centre at finite $\lambda$, where $\theta\simeq-2/(\lambda_c-\lambda)$ diverges in a caustic, exactly as for supercritical observers and for Maxwell rays [Eq.~\eqref{eq:null-lambda}]. For $E<\mu_0$ the rays are reflected at $r_t$, Eq.~\eqref{eq:rt}, where $P_E\to0$, $D_E\to1$ and $\lambda_t-\lambda\simeq1/|\theta|$: a fold caustic at finite $\lambda$. In the core the affine parameter follows in closed form from $\dd\lambda=-\dd r/P_E$,
\begin{equation}
\Delta\lambda_E\simeq
\begin{cases}
\dfrac{1}{\sigma}\,\operatorname{arcsinh}\!\left[\dfrac{\sigma r_1}{\sqrt{E^{2}-\mu_0^{2}}}\right], & E>\mu_0,\\[2ex]
\dfrac{1}{\sigma}\,\ln\dfrac{r_1}{r}\;\longrightarrow\;\infty, & E=\mu_0,\\[2ex]
\dfrac{1}{\sigma}\,\operatorname{arccosh}\!\left[\dfrac{\sigma r_1}{\sqrt{\mu_0^{2}-E^{2}}}\right], & E<\mu_0 ,
\end{cases}
\label{eq:lambda-E}
\end{equation}
the counterpart of Eq.~\eqref{eq:tau-eps}. 

\subsection{Completeness}
\label{sec:photon-complete}

Completeness is judged with $\lambda$, the parameter of the Hamiltonian~\eqref{eq:H}, which reduces to the affine parameter of a null geodesic where $\mu\to0$. We consider the idealised static metric with its smooth maximal extension stipulated; this is independent of the dynamical stability of its inner Cauchy horizon, and global statements additionally require regular horizon charts and control of all ends of the chosen extension. Since $f$ and $\mu^{2}$ are smooth functions of $r^{2}$, the Hamiltonian~\eqref{eq:H} is smooth at the origin in Cartesian-like coordinates, and $r=0$ is an ordinary point: a ray reaching it continues antipodally, with no extension to $r<0$ needed. The comparison with the signed-radius analytic extension of Hayward is understood as specified in Sec.~\ref{sec:continuation}; since $f\mu^{2}$ is even and analytic, the argument of Sec.~\ref{sec:continuation} applies unchanged. Horizon crossing is described in regular ingoing or outgoing coordinates, and rays with $J\neq0$ are turned back by the centrifugal barrier. The three radial classes then complete as follows.

(1) Supercritical, $E>\mu_0$. The ray passes through the centre at finite $\lambda$, continues antipodally, crosses the core, the inner horizon and the shell, and enters a new exterior region. Since $f\mu^{2}\leq\mu_0^{2}<E^{2}$ everywhere, no barrier stops it, and with $U_{\min}<0$ the minimum of $f\mu^{2}$, attained between the horizons,
\begin{equation}
\lambda(R)-\lambda_c\;\geq\;\frac{R}{\sqrt{E^{2}-U_{\min}}}\;\xrightarrow{\;R\to\infty\;}\;\infty ,
\label{eq:lambda-escape}
\end{equation}
the counterpart of Eq.~\eqref{eq:tau-escape}.

(2) Subcritical, $0<E<\mu_0$. For a ray entering the core, reflection at the inner simple turning point $r_t$ occurs at finite $\lambda$, after which the ray moves outward through the inner horizon into a new exterior region. The decay $\mu\propto r^{-5/2}$ makes $U=f\mu^2\to0$ at infinity and permits escape when the energy exceeds the exterior potential barrier. Let $U_m=U(r_m)$ denote the exterior maximum. If $E^2>U_m$, the outgoing ray reaches infinity only as $\lambda\to\infty$. If $E^2<U_m$, a core-connected ray is reflected at the exterior barrier and cycles through successive regions, with a finite, nonzero parameter interval per cycle. Rays arriving from infinity at these lower energies instead reflect on the outer side of the exterior barrier without entering the core.

At the exceptional energy $E^2=U_m$, the maximum is a double root of the radial equation. Since $U''(r_m)<0$,
\begin{equation}
E^2-U(r)=-\frac{1}{2}U''(r_m)(r-r_m)^2
+O\!\left((r-r_m)^3\right),
\label{eq:outer-separatrix}
\end{equation}
so $\int\dd r/\sqrt{E^2-U}$ diverges logarithmically. A separatrix ray approaches the unstable stationary orbit only at infinite $\lambda$; the stationary orbit itself extends for all $\lambda$. Thus the equality case is complete as well, but is not a finite-time reflection or cycle.

(3) Critical, $E=\mu_0$. Here $\lambda\to\infty$ is attained at the centre itself, by Eq.~\eqref{eq:lambda-crit}.

The past extension is obtained by the time-reversed classification: scattering rays extend to infinity, trapped rays undergo indefinitely many cycles, separatrices approach stationary orbits, and a critical ray may emerge asymptotically from the centre. It is therefore not necessary that every ray originate at infinity. For the positive-energy radial Hamiltonian trajectories considered here, neither time direction has a finite-$\lambda$ endpoint at the regular centre, a regular horizon, an asymptotic infinity or a stationary-orbit limit on the stipulated maximal extension.

One caveat concerns the choice of parameter. Along an outgoing ray $\dd r/\dd\lambda\to E$ while $\mu\propto r^{-5/2}$, so the ray reaches infinity at infinite $\lambda$ but at finite $\tau=\int\mu\,\dd\lambda$ and finite $\tilde\tau=\int\mu^{2}\,\dd\lambda$. In particular, the conformal metric~\eqref{eq:gtilde} is timelike-incomplete at infinity, because its conformal factor vanishes there. This is not an incompleteness of the photon dynamics: the photon becomes massless at large $r$, and neither $\tau$ nor $\tilde\tau$ is then a useful clock. These statements do not establish completeness of a perturbed spacetime or exclude mass-inflation singularities.

Table~\ref{tab:photons} compares the three cases with Maxwell rays and with Schwarzschild spacetime. At $\kappa=0$ the potential vanishes, every radial ray behaves like a supercritical one, and the single complete class of Sec.~\ref{sec:maxwell} is recovered. In a Schwarzschild black hole surrounded by a plasma, by contrast, Eq.~\eqref{eq:radial} with $f=1-2M/r$ drives every ray that crosses the horizon to $r=0$ at finite $\lambda$, where it ends: the photon dynamics is incomplete for massive and massless rays alike. Completeness of the photon dynamics is therefore inherited from the T-duality geometry. What the London mass adds is the threshold $\mu_0$, which separates rays that pass through the centre, rays that are turned back and critical rays that approach it for all $\lambda$, together with the medium defocusing in Eq.~\eqref{eq:cancel}.

\begin{table*}[t]
\caption{Radial photon rays near the centre of the T-duality black hole. $D$ is the right-hand side of Eq.~\eqref{eq:inv-theta-photon} at the centre (at $r_t$ for $E<\mu_0$); the observer counterparts are listed in Table~\ref{tab:observers}.}
\label{tab:photons}
\begin{ruledtabular}
\begin{tabular}{lcccc}
& $D$ & $\theta$ & $\lambda$ to $r=0$ or $r_t$ & Complete \\
\hline
Maxwell, $\kappa=0$ & $1/2$ & $-\infty$ (caustic) & finite & yes \\
$E>\mu_0$ & $1/2$ & $-\infty$ (caustic) & finite & yes \\
$E=\mu_0$ & $0$ & $-3\sigma$ & $\infty$ & yes \\
$E<\mu_0$ & $1$ & $-\infty$ (fold) & finite & yes \\
Schwarzschild & --- & $-\infty$ (singular) & finite & no \\
\end{tabular}
\end{ruledtabular}
\end{table*}

\section{Conclusions}
\label{sec:conclusions}

We have analysed geodesic completeness in the regular core of the T-duality black hole with the Raychaudhuri equation, first for massive observers and Maxwell rays, and then for photons that acquire a mass in a superconducting core.

For radial observers of arbitrary specific energy $\varepsilon$, the Raychaudhuri equation gives the exact evolution of the inverse expansion, Eq.~\eqref{eq:inverse-time-integral}, and, combined with the radial geodesic equation, the exact proper-time quadrature~\eqref{eq:JN21-eps}; closed forms follow in the leading de Sitter approximation, and no constant-$K_\varepsilon$ assumption is needed. The behaviour at the centre is fixed by the value there of $D_\varepsilon=\tfrac13+K_\varepsilon$. Critical observers, $\varepsilon=1$, with $D_\varepsilon\to0$, keep a finite expansion and approach the centre only after a logarithmically divergent proper time, $\tau\simeq\LdS\ln(r_0/r)$, as found in Ref.~\cite{Jusufi:2024}; we have obtained this proper time in closed form, Eq.~\eqref{eq:tau-critical-closed}, and shown that the Raychaudhuri and geodesic routes to it coincide exactly through the identity~\eqref{eq:D-identity}; this behaviour is special to the critical orbit, an unstable equilibrium. Supercritical observers, with $D_\varepsilon\to1/2$, reach the centre at finite proper time through a caustic and pass through it; subcritical observers, with $D_\varepsilon\to1$, are reflected at a fold caustic and cycle through successive regions of the maximal extension. All three classes are complete. Where geodesics reach the centre at finite proper time, their continuation is unique and regular because the lapse is analytic and even in $r$, so that $r<0$ is the antipodal part of the same spacetime; by contrast, the signed-radius analytic extension of the Hayward metric is incomplete~\cite{Zhou:2022}. In Schwarzschild spacetime the $\varepsilon=1$ congruence obeys $1/\Theta_1-1/\Theta_0=\tau$ exactly, and for every $\varepsilon$ the infalling geodesics reach the curvature singularity at finite proper time, where they end. The T-duality core does not prevent the expansion from diverging. It turns the endpoint into a regular point, makes gravity repulsive near the centre, and thereby replaces the single, incomplete class of Schwarzschild infall by three complete classes.

Radial Maxwell rays have the flat-space expansion $\theta=-2E/r$, pass through the regular centre and are complete, but form a single class, since a massless ray carries no energy scale and the Ricci coupling of the wave operator leaves its rays null. We supply a dispersive scale by assigning a London response to the effective core source itself. Its background density and pressures are already included in the metric, and are counted once. The background is neutral, with zero electromagnetic field and current, while its stiffness is nonzero. A symmetry-breaking scalar potential with vacuum-energy offset $\rho_0$ provides a homogeneous central example: $V(v)=\rho_0$ supplies the de Sitter stress and $\mu_0^2=4\pi q^2v^2/\hbar^2$ supplies the photon mass. This example is exact at the centre and approximate throughout the inner core, with corrections of relative order $r^{2}/\lz^{2}$, but no static scalar theory can reconstruct the full radial source or stiffness profile, because it forces $p_\perp=-\rho$.

Under the constitutive assumption $\mu^2=4\pi\kappa\rho$, the photon mass tracks the local curvature, $\mu_0^{2}=\kappa\Leff/2$, and $\lambda_L(0)/\LdS=\sqrt{2/(3\kappa)}$ for every de Sitter core obeying the same relation. In the core wave operator the London term appears alongside the Ricci coupling, with equal coefficients at $\kappa=2$, but only the London term survives in geometric optics. Photon rays then follow, up to reparametrisation, timelike geodesics of $\mu^{2}g_{\mu\nu}$, so the null--timelike correspondence suggested in Ref.~\cite{Jusufi:2024} holds at the level of rays. The photon Raychaudhuri equation reproduces the classification of observers with $\varepsilon=1$ replaced by $E=\mu_0$ and $1/\LdS$ by $\sigma$: quanta above the central plasma frequency cross the centre through a caustic, quanta below it are turned back, and critical quanta approach the centre asymptotically with finite expansion and vanishing focusing rate. The medium adds a density-gradient term that defocuses near the centre. All three classes are complete in the Hamiltonian parameter $\lambda$, so the mass enriches the phenomenology of light in the core without being needed for completeness, which is inherited from the regular geometry.

The analysis rests on three approximations: the prescribed-response model~\eqref{eq:Sresp} on a fixed background; the homogeneous-medium approximation for the core wave equation, valid for $\kappa M\gg\lz$; and the massive-particle eikonal approximation, whose near-threshold core regime requires $\kappa\gg1$ and which fails near turning points and at the critical endpoint. The energy classes describe Hamiltonian rays; at the wave level the angular or longitudinal derivative terms and the regular central boundary condition must be retained, and no universal transmission threshold has been derived. The completeness statements refer to the idealised maximal extension, without a claim of stability under inner-horizon perturbations. A canonical potential alone cannot reproduce the full source, because it cancels from $\rho_\psi+p_{r,\psi}$ in Eq.~\eqref{eq:NECpsi}, whereas the radial and tangential pressures require the nontrivial anisotropy~\eqref{eq:source-anisotropy}. A microscopic completion of the constitutive profile remains open. The local potential construction does not derive the density-gradient contribution to the ray force; that contribution follows from the prescribed full profile. For $\lz$ near $l_{\rm Pl}$ and large $M$, the central curvature is trans-Planckian, so the effective description is strained near the centre.

Several directions follow naturally: reconstructing a microscopic source action that reproduces both the full anisotropic stress and the electromagnetic stiffness; testing its stability; deriving $\kappa$ from the string states that set $\lz$; studying the polar Proca sector and quasinormal modes; and a dynamical analysis of the inner horizon.


\begin{thebibliography}{99}

\bibitem{Borissova:2025}
J.~Borissova, S.~Liberati, and M.~Visser, \href{https://doi.org/10.1103/PhysRevD.111.104054}{Phys.\ Rev.\ D \textbf{111}, 104054 (2025)}.

\bibitem{Penrose:1965}
R.~Penrose, \href{https://doi.org/10.1103/PhysRevLett.14.57}{Phys.\ Rev.\ Lett.\ \textbf{14}, 57 (1965)}.

\bibitem{Hawking:1970}
S.~W.~Hawking and R.~Penrose, \href{https://doi.org/10.1098/rspa.1970.0021}{Proc.\ R.\ Soc.\ Lond.\ A \textbf{314}, 529 (1970)}.

\bibitem{HawkingEllis}
S.~W.~Hawking and G.~F.~R.~Ellis, \href{https://doi.org/10.1017/CBO9780511524646}{\emph{The Large Scale Structure of Space-Time}} (Cambridge University Press, Cambridge, 1973).

\bibitem{Bardeen:1968}
J.~M.~Bardeen, in \emph{Proceedings of GR5} (Tbilisi, 1968), p.~174.

\bibitem{Dymnikova:1992}
I.~Dymnikova, \href{https://doi.org/10.1007/BF00760226}{Gen.\ Relativ.\ Gravit.\ \textbf{24}, 235 (1992)}.

\bibitem{AyonBeato:2000}
E.~Ay\'on-Beato and A.~Garc\'ia, \href{https://doi.org/10.1016/S0370-2693(00)01125-4}{Phys.\ Lett.\ B \textbf{493}, 149 (2000)}, \href{https://arxiv.org/abs/gr-qc/0009077}{arXiv:gr-qc/0009077}.

\bibitem{Hayward:2006}
S.~A.~Hayward, \href{https://doi.org/10.1103/PhysRevLett.96.031103}{Phys.\ Rev.\ Lett.\ \textbf{96}, 031103 (2006)}, \href{https://arxiv.org/abs/gr-qc/0506126}{arXiv:gr-qc/0506126}.

\bibitem{Nicolini:2006}
P.~Nicolini, A.~Smailagic, and E.~Spallucci, \href{https://doi.org/10.1016/j.physletb.2005.11.004}{Phys.\ Lett.\ B \textbf{632}, 547 (2006)}, \href{https://arxiv.org/abs/gr-qc/0510112}{arXiv:gr-qc/0510112}.

\bibitem{Poisson:1990}
E.~Poisson and W.~Israel, \href{https://doi.org/10.1103/PhysRevD.41.1796}{Phys.\ Rev.\ D \textbf{41}, 1796 (1990)}.

\bibitem{CarballoRubio:2018}
R.~Carballo-Rubio, F.~Di~Filippo, S.~Liberati, C.~Pacilio, and M.~Visser, \href{https://doi.org/10.1007/JHEP07(2018)023}{J.\ High Energy Phys.\ \textbf{07} (2018) 023}, \href{https://arxiv.org/abs/1805.02675}{arXiv:1805.02675}.

\bibitem{Nicolini:2019}
P.~Nicolini, E.~Spallucci, and M.~F.~Wondrak, \href{https://doi.org/10.1016/j.physletb.2019.134888}{Phys.\ Lett.\ B \textbf{797}, 134888 (2019)}.

\bibitem{Padmanabhan:1997}
T.~Padmanabhan, \href{https://doi.org/10.1103/PhysRevLett.78.1854}{Phys.\ Rev.\ Lett.\ \textbf{78}, 1854 (1997)}.

\bibitem{Smailagic:2003}
A.~Smailagic, E.~Spallucci, and T.~Padmanabhan, \href{https://arxiv.org/abs/hep-th/0308122}{arXiv:hep-th/0308122}.

\bibitem{Fontanini:2006}
M.~Fontanini, E.~Spallucci, and T.~Padmanabhan, \href{https://doi.org/10.1016/j.physletb.2005.12.039}{Phys.\ Lett.\ B \textbf{633}, 627 (2006)}, \href{https://arxiv.org/abs/hep-th/0509090}{arXiv:hep-th/0509090}.

\bibitem{Gaete:2022ukm}
P.~Gaete, K.~Jusufi, and P.~Nicolini, \href{https://doi.org/10.1016/j.physletb.2022.137546}{Phys.\ Lett.\ B \textbf{835}, 137546 (2022)}, \href{https://arxiv.org/abs/2205.15441}{arXiv:2205.15441}.

\bibitem{Chen:2026nna}
C.~M.~Chen, K.~Jusufi, and D.~Singleton, \href{https://arxiv.org/abs/2609.18528}{arXiv:2609.18528}.

\bibitem{Zhou:2022}
T.~Zhou and L.~Modesto, \href{https://doi.org/10.1103/PhysRevD.107.044016}{Phys.\ Rev.\ D \textbf{107}, 044016 (2023)}, \href{https://arxiv.org/abs/2208.02557}{arXiv:2208.02557}.

\bibitem{Jusufi:2024}
K.~Jusufi and P.~Nicolini, \href{https://doi.org/10.1140/epjc/s10052-025-15018-8}{Eur.\ Phys.\ J.\ C \textbf{85}, 1291 (2025)}, \href{https://arxiv.org/abs/2410.19613}{arXiv:2410.19613}.

\bibitem{Raychaudhuri:1955}
A.~Raychaudhuri, \href{https://doi.org/10.1103/PhysRev.98.1123}{Phys.\ Rev.\ \textbf{98}, 1123 (1955)}.

\bibitem{London:1935}
F.~London and H.~London, \href{https://doi.org/10.1098/rspa.1935.0048}{Proc.\ R.\ Soc.\ Lond.\ A \textbf{149}, 71 (1935)}.

\bibitem{Ginzburg:1950}
V.~L.~Ginzburg and L.~D.~Landau, Zh.\ Eksp.\ Teor.\ Fiz.\ \textbf{20}, 1064 (1950).

\bibitem{Weinberg:1986}
S.~Weinberg, \href{https://doi.org/10.1143/PTPS.86.43}{Prog.\ Theor.\ Phys.\ Suppl.\ \textbf{86}, 43 (1986)}.

\bibitem{Anderson:1963}
P.~W.~Anderson, \href{https://doi.org/10.1103/PhysRev.130.439}{Phys.\ Rev.\ \textbf{130}, 439 (1963)}.

\bibitem{Higgs:1964}
P.~W.~Higgs, \href{https://doi.org/10.1103/PhysRevLett.13.508}{Phys.\ Rev.\ Lett.\ \textbf{13}, 508 (1964)}.

\bibitem{Synge:1960}
J.~L.~Synge, \emph{Relativity: The General Theory} (North-Holland, Amsterdam, 1960).

\bibitem{Perlick:2017}
V.~Perlick and O.~Yu.~Tsupko, \href{https://doi.org/10.1103/PhysRevD.95.104003}{Phys.\ Rev.\ D \textbf{95}, 104003 (2017)}.

\bibitem{Perlick:2022}
V.~Perlick and O.~Yu.~Tsupko, \href{https://doi.org/10.1016/j.physrep.2021.10.004}{Phys.\ Rep.\ \textbf{947}, 1 (2022)}, \href{https://arxiv.org/abs/2105.07101}{arXiv:2105.07101}.

\bibitem{Bhattacharjee:2019}
C.~Bhattacharjee, J.~C.~Feng, and S.~M.~Mahajan, \href{https://doi.org/10.1103/PhysRevD.99.024027}{Phys.\ Rev.\ D \textbf{99}, 024027 (2019)}, \href{https://arxiv.org/abs/1903.08117}{arXiv:1903.08117}.

\bibitem{Turimov:2026}
B.~Turimov \emph{et al.}, \href{https://arxiv.org/abs/2610.03761}{arXiv:2610.03761}.

\bibitem{Rosa:2012}
J.~G.~Rosa and S.~R.~Dolan, \href{https://doi.org/10.1103/PhysRevD.85.044043}{Phys.\ Rev.\ D \textbf{85}, 044043 (2012)}.


\bibitem{JusufiLobo:2026dressing}
K.~Jusufi and F.~S.~N.~Lobo, ``Dressing a naked singularity with zero-point length,'' \href{https://arxiv.org/abs/2610.24000}{arXiv:2610.24000 [gr-qc]} (2026).

\bibitem{JusufiLobo:2026neutron}
K.~Jusufi and F.~S.~N.~Lobo, ``Planck-scale black hole remnants inside neutron stars,'' \href{https://arxiv.org/abs/2610.10589}{arXiv:2610.10589 [gr-qc]} (2026).


\bibitem{AnandJusufiBambi:2026topology}
A.~Anand, K.~Jusufi, and C.~Bambi, ``Zero-point length as a topological protection of black hole regularity,'' \href{https://arxiv.org/abs/2603.02295}{arXiv:2603.02295 [gr-qc]} (2026).

\bibitem{Bronnikov:2001}
K.~A.~Bronnikov, ``Spherically symmetric false vacuum: No-go theorems and global structure,'' \href{https://doi.org/10.1103/PhysRevD.64.064013}{Phys.\ Rev.\ D \textbf{64}, 064013 (2001)}, \href{https://arxiv.org/abs/gr-qc/0104092}{arXiv:gr-qc/0104092}.


\end{thebibliography}
\end{document}